\documentclass[journal,10pt]{IEEEtran}
\usepackage[english]{babel}
\usepackage[utf8]{inputenc}
\usepackage{amsmath}
\usepackage{amssymb}
\usepackage{bbm}

\usepackage{algorithm,algcompatible,amsmath}
\newtheorem{theorem}{Theorem}[section]

\newtheorem{lemma}[theorem]{Lemma}
\usepackage{graphicx}
\usepackage{amstext}
\usepackage{amsmath}
\usepackage{amsfonts,amssymb}
\usepackage{amsmath,tabu}
\usepackage{amssymb}
\usepackage[dvips]{epsfig}
\usepackage{epsfig}
\usepackage[dvips]{graphicx}
\usepackage{multirow}
\usepackage[utf8]{inputenc}
\usepackage[english]{babel}
\usepackage{caption}
\usepackage{float}
\usepackage[nospace,noadjust]{cite}
\usepackage{bm}
\usepackage{xcolor,colortbl}

\ifCLASSOPTIONcompsoc
\usepackage[caption=false, font=normalsize, labelfont=sf, textfont=sf, subrefformat=parens,labelformat=parens]{subfig}
\else
\usepackage[caption=false, font=footnotesize, subrefformat=parens,labelformat=parens]{subfig}
\fi

\def\bth{{\boldsymbol{\theta}}}

\def\bPsi{{\boldsymbol{\Psi}}}
\def\bLam{{\boldsymbol{\Lambda}}}
\def\blam{{\boldsymbol{\lambda}}}
\def\b1{{\boldsymbol{1}}}
\def\c1{{\textcircled{a}}}

\def\bn{{\boldsymbol{n}}}

\def\bs{{\boldsymbol{s}}}

\def\bw{{\boldsymbol{w}}}
\def\bx{{\boldsymbol{x}}}
\def\by{{\mathbf{y}}}

\def\bA{{\mathbf{A}}}
\def\bB{{\boldsymbol{B}}}
\def\bC{{\boldsymbol{C}}}
\def\bD{{\boldsymbol{D}}}
\def\bE{{\boldsymbol{E}}}
\def\bF{{\mathbf{F}}}
\def\bG{{\boldsymbol{G}}}

\def\bI{{\mathbf{I}}}
\def\bJ{{\boldsymbol{J}}}

\def\bP{{\mathbf{P}}}
\def\bQ{{\boldsymbol{Q}}}
\def\bR{{\mathbf{R}}}

\def\bT{{\mathbf{T}}}
\def\bU{{\boldsymbol{U}}}
\def\bV{{\boldsymbol{V}}}

\def\bX{{\boldsymbol{X}}}
\def\bY{{\boldsymbol{Y}}}
\def\bZ{{\boldsymbol{Z}}}

\def\bzero{{\boldsymbol{0}}}

\newcommand{\trace}{\textrm{Tr}}

\providecommand{\keywords}[1]{\textbf{\textit{Index terms---}} #1}
\makeatletter
\def\thanks#1{\protected@xdef\@thanks{\@thanks
        \protect\footnotetext{#1}}}
\makeatother

\usepackage{color}

\newcommand{\red}{\textcolor{black}}

\allowdisplaybreaks[4]

\definecolor{Gray}{gray}{0.85}
\definecolor{White}{gray}{1}
\newcolumntype{a}{>{\columncolor{Gray}}c}

\begin{document}

\title{New Methods for MLE of Toeplitz Structured Covariance Matrices with Applications to RADAR Problems}

\author{Augusto Aubry,~\IEEEmembership{Senior Member,~IEEE}, Prabhu Babu, Antonio De Maio,~\IEEEmembership{Fellow,~IEEE}, and Massimo Rosamilia,~\IEEEmembership{Member,~IEEE}
\thanks{A. Aubry and A. De Maio  are with the Department of Electrical Engineering and Information Technology, Universita degli Studi di Napoli ``Federico II”, DIETI, Via Claudio 21, I-80125 Napoli, Italy (E-mail: augusto.aubry@unina.it, ademaio@unina.it).}
\thanks{P. Babu is with CARE, IIT Delhi, New Delhi, 110016, India (E-mail: prabhubabu@care.iitd.ac.in)
}
\thanks{M. Rosamilia is with the National Inter-University Consortium for Telecommunications, 43124 Parma, Italy (e-mail: massimo.rosamilia@unina.it).}
}
\markboth{Submitted to IEEE Trans. on Signal Processing}{...}
\maketitle

\begin{abstract}
This work considers Maximum Likelihood Estimation (MLE) of a Toeplitz structured covariance matrix. In this regard, an equivalent reformulation of the MLE problem is introduced and two iterative algorithms are proposed for the optimization of the equivalent statistical learning framework. Both the strategies are based on the Majorization Minimization (MM) paradigm and hence enjoy nice properties such as monotonicity and ensured convergence to a stationary point of the equivalent MLE problem. The proposed framework is also extended to deal with MLE of other practically relevant covariance structures, namely, the banded Toeplitz, block Toeplitz, and Toeplitz-block-Toeplitz. Through numerical simulations, it is shown that the new methods provide excellent performance levels in terms of both mean square estimation error (which is very close to the benchmark Cramér-Rao Bound (CRB)) and signal-to-interference-plus-noise ratio, especially in comparison with state of the art strategies.

\end{abstract}
\keywords{Toeplitz covariance matrix, Maximum likelihood estimation, Banded Toeplitz, {Block-Toeplitz,} Toeplitz-block-Toeplitz, Adaptive radar signal processing, Array processing, Spectral estimation}
\section{Introduction}
Estimation of the {data} covariance matrix has diverse applications in radar signal processing, such as direction
of arrival estimation, target detection, adaptive beamforming, and sidelobe canceller design \cite{amf, stoica,app,farina}. In these situations, the interference covariance matrix is estimated from the secondary/training data, which are assumed target-free and collected from spatial and/or temporal returns corresponding to range cells close to the one of interest. {When the data follows a complex, {zero-mean,} circular Gaussian distribution, it is well known that} the Sample Covariance Matrix (SCM) is the unstructured {Maximum Likelihood (ML) estimate} {of the covariance matrix. However, in the presence of a small number of training data and/or when mismatches in training data spectral properties occur,} it does not always represent a {reliable choice for} the covariance {inference}~\cite{scm1,scm2}. A well-known strategy, often discussed in the open literature to improve the performance of a covariance estimator, relies on the incorporation of some \emph{a priori} knowledge about its {underlying} structure. For instance, in some radar applications, it is customary to suppose that data come from a  stationary Gaussian random process, leading to a {Hermitian symmetric} Toeplitz Structured Covariance (TSC) matrix. Leveraging this information, one can obtain (under the design conditions) a more reliable estimator {than} the SCM \cite{toep}. {Aside} radar applications, the estimation of a TSC matrix is encountered in speech recognition \cite{speech}, spectral estimation \cite{stoica}, {gridless compressive sensing~\cite{6857424, 6891350,6780606}}, and hyperspectral imaging \cite{imaging}.

So far, several algorithms have been proposed for estimating a TSC matrix. Let us first discuss those for {ML Estimation (MLE)}. According to the Caratheodory parametrization \cite{stoica,caratheodoryp1,caratheodoryp2} a Toeplitz covariance matrix $\bT \in \mathbb{H}^{m \times m}$ can always be decomposed as\footnote{{Notice that the parametrization is unique provided that the rank of $\bT<m$~\cite{yang2018sparse}.}}
\begin{equation}\label{cp}
    \begin{array}{ll}
    \bT = {\bA}\tilde{\bP}{\bA}^{H}; \:\: [\tilde{\bP}]_{k,k}\geq 0
    \end{array},
\end{equation}
where
\begin{equation}
\bA=
\begin{bmatrix}
1 & \cdots & 1\\
e^{j\omega_{1}}  & \cdots & e^{j\omega_{r}} \\
\vdots  & \ddots & \vdots  \\
e^{j(m-1)\omega_{1}}  & \cdots & e^{j(m-1)\omega_{r}}
\end{bmatrix},
\tilde{\bP} =
  {\begin{bmatrix}
    \tilde{p}_{1} &\dots&0 \\
     \vdots & \ddots & \vdots \\
    0&\dots& \tilde{p}_{r}
  \end{bmatrix}},
\end{equation}
$\omega_{i}$ and $\tilde{p}_{i}$, $i=1,2, \cdots,r \leq m$, denote some angular frequencies and their corresponding powers {while} $r$ indicates the rank of $\bT$. Capitalizing on this parametrization, Circulant Embedding (CE) of Toeplitz matrix (\cite{ce1,ce2,ce3}) can be used  to compute approximately the ML estimate of $\bT$. According to CE, a Positive SemiDefinite {(PSD)} $m \times m$ Toeplitz matrix is modeled as
\begin{equation}\label{ce}
    \begin{array}{ll}
    \bT = \tilde{\bF}\bP\tilde{\bF}^{H}; \:\: \bP= {\textrm{diag}}([p_{1},p_{2},\cdots,p_{L}]), p_{k} \geq 0 ,
    \end{array}
\end{equation}
where $\tilde{\bF} = [\bI_{m \times m} \: \bzero_{m \times (L-m)}]\bF$, $\bI_{m \times m}$ is the identity matrix of size $m \times m$, $\bzero_{m \times L-m}$ is the zero matrix of size $m \times L-m$, $\bF$ is the normalized Discrete Fourier Transform (DFT) matrix of size $L \geq 2m-1$ and $\bP$ is a diagonal matrix of size $L \times L$ with diagonal elements $p_{k} \geq 0$. Therefore, the matrix $\bT$ is completely parameterized by the diagonal matrix $\bP$. Although estimating the Toeplitz covariance matrix using CE seems attractive, the representation in (\ref{ce}) is valid only for a subset of Toeplitz covariance matrices. This can be intuitively justified because the Caratheodory parametrization in (\ref{cp}) does not give restrictions on the frequencies spacing, while the CE in (\ref{ce}) strictly requires the frequencies to lie on the Fourier grid. Hence, for some Toeplitz matrices, the parametrization in (\ref{ce}) is only {approximated}. Based on CE, \cite{em1} and \cite{em2} {have} proposed an iterative algorithm based on Expectation-Maximization (EM) for MLE of $\bT$. By modifying the M step in the EM procedure, in \cite{band} the technique has been extended to deal with the banded Toeplitz covariance case. In \cite{melt}, still leveraging CE framework, a Majorization Minimization (MM) based optimization, with faster convergence than the EM {of} \cite{em1} and \cite{em2}, has been introduced. In \cite{ip} a closed-form estimator {has been} designed by invoking the extended invariance principle to deal with the Toeplitz constraint. Finally, in \cite{eastr}, an efficient approximation of a Toeplitz covariance matrix under a rank constraint has been {handled} forcing the eigenvectors to be the same as those of the SCM whereas the Toeplitz constraint has been explicitly imposed while estimating the eigenvalues. Other than the MLE, several other alternative paradigms have been considered for the problem at hand. Recently, in \cite{Augusto} the Toeplitz structure is forced together with a condition number constraint via SCM projection onto a suitable constraint set.  Other geometric based approaches for the TSC {estimation} have also been proposed in \cite{geo1,geo2}.

In this manuscript, two iterative algorithms {referred to} as \textbf{A}lternating Projection Based \textbf{TO}eplitz Covariance \textbf{M}atrix Estimation 1 (ATOM1)  and ATOM2 are devised leveraging a suitable reformulation of the MLE problem and the MM framework. Both ATOM1 and ATOM2 involve the construction of a {bespoke} {surrogate function (s.f.)} {along} with its optimization. Specifically, the two procedures construct distinct s.f. and therefore solve different surrogate minimization problems. While ATOM1 addresses the surrogate minimization problem using the Alternating Direction Method of Multipliers (ADMM), ATOM2 handles it {either} via alternating projection or Dykstra's algorithm. However, both the procedures directly estimate the Toeplitz covariance matrix without {forcing a} reparametrization via the CE. ATOM2 is also extended to include other constraints, such as banded Toeplitz, {block-Toeplitz,} and Toeplitz-block-Toeplitz structures. The major contributions of this paper can be summarized as follows:
\begin{enumerate}
    \item Two iterative algorithms ATOM1 and ATOM2 are proposed based on the MM framework to address MLE of a Toeplitz covariance matrix. Their computational complexities are thoroughly discussed. Also, the convergence of the procedures to a stationary point of the {equivalent} MLE problem is established.
    \item The extensions of ATOM2 to handle additional covariance structures, such as banded Toeplitz, {block-Toeplitz, and} Toeplitz-block-Toeplitz.
    \item The {derivation of the Cramér-Rao Bound (CRB)} for the estimation of Toeplitz, banded Toeplitz, and Toeplitz-block-Toeplitz covariance matrices are {provided}.
    \item {Performance comparisons of the} proposed algorithms {(included} their extensions{)} with some state-of-the-art procedures via numerical simulations {are illustrated,} using the Mean Square Error (MSE) and {the} Signal-to-Interference-plus-Noise Ratio (SINR) (for case studies related to radar {applications}) as performance metrics.
\end{enumerate}
The organization of the paper is as follows. The MLE problem of Toeplitz covariance matrix for complex, {zero-mean,} circular Gaussian observations is formulated in Section \ref{sec:2}. In Section \ref{sec:3}, ATOM1 and ATOM2 algorithms are proposed, along with a discussion on their computational complexity and implementation aspects. Also, their convergence properties are studied. At the end of {this} section, the extension of ATOM2 to handle additional constraints along with the Toeplitz requirement is discussed too. In Section \ref{sec:crlb}, the {CRB} for the {estimation} of Toeplitz, banded Toeplitz, and Toeplitz-block-Toeplitz covariance matrices {is} computed. In Section \ref{sec:4}, the proposed algorithms are compared with some state-of-the-art {techniques}, and finally, concluding remarks are given in Section \ref{sec:5}.

\subsection{Notation}
Throughout the paper, {{bold}} capital and {{bold}} small letter denote matrix and vector, respectively. A scalar is represented by a small letter. The value taken by an optimization vector ${\bx}$ at the ${t^{th}}$ iteration is denoted by ${\bx_{t}}$. 
{Furthermore, $\mathbb{R}$ is used to denote the set of real numbers,  $\mathbb{R}^{m}$ and $\mathbb{C}^{m}$ are used to represent the sets of $m$ dimensional vectors of real and complex numbers, respectively, whereas $\mathbb{R}^{m \times m}$, $\mathbb{C}^{m \times m}$, and
$\mathbb{H}^{m \times m}$ are used to represent the sets of $m \times m$ matrices of real numbers, $m \times m$ matrices of complex numbers, and $m \times m$ Hermitian matrices, respectively}. Superscripts $(\cdot)^{T}$, $(\cdot)^{*}$,  $(\cdot)^{H}$, and $(\cdot)^{-1}$ indicate the transpose, complex conjugate, complex conjugate transpose, and inverse, respectively. {For any $x \in \mathbb{R}$, $\lceil x \rceil$ returns the least integer greater than or equal to $x$}. The trace and the determinant of a matrix $\bX$ are denoted by ${\rm{Tr}}(\bX)$ and $|\bX|$, respectively. The notation $[{\bX}]_{i}$ is used to represent the $i^{th}$ column of the matrix $\bX$. The {symbol} $\otimes$ indicates the Kronecker product while the gradient of a function $f$ is denoted by $\nabla f$. The {symbol} $\succeq$ (and its strict form $\succ$) is used to denote the generalized matrix inequality: for any $\bX \in \mathbb{H}^{m \times m}$,  $\bX \succeq  0$ means that $\bX$ is a {PSD} matrix ($\bX \succ 0$ for positive definiteness). Besides, for any $\bX \in \mathbb{H}^{m \times m}$, ${\rm{eig}}(\bX)$ is the vector collecting the eigenvalues of $\bX$ {(sorted in increasing order)}. The Euclidean norm of the vector $\bx$ is denoted by ${\|\bx\|}_{2}$, $|\bx|$ indicates the element wise modulus of the vector $\bx$. The notation $\textbf{E}[\cdot]$ stands for statistical expectation. Finally, for any $\bX,\bY \in \mathbb{R}^{m\times m}$, $\max(\bX,\bY)$ refers to the matrix containing the element wise maximum between  $\bX$ and $\bY$.

\section{Problem Formulation}\label{sec:2}
Let us assume the availability of $n$ independent and identically distributed vectors $\{\by_{1}, \by_{2}, \cdots,\by_{n}\}$, where each $\by_{i}$ is of size $m$ and follows a $m$-variate complex, {zero-mean,} circular Gaussian distribution with covariance matrix ${\bR \succ0}$. The maximum likelihood covariance estimation problem can be formulated as
\begin{equation} \label{eq:11}
\begin{array}{ll}
\underset{\bR \succ 0}{\rm minimize} \: {\bar{f}}(\bR) {=}\dfrac{1}{n}\displaystyle\sum_{i=1}^{n}\by_{i}^{H}\bR^{-1}\by_{i} + \log|\bR| .
\end{array}
\end{equation}
If $n \geq m$, Problem (\ref{eq:11}) has a unique minimizer with probability one which is given by the SCM, i.e., $\bR_{\textrm{SCM}} = \dfrac{1}{n}\displaystyle\sum_{i=1}^{n}\by_{i}\by_{i}^{H}$.  However, if the random process, {where} each observation is drawn, is  stationary (at least in wide sense) then the covariance matrix also exhibits a Toeplitz structure which can be capitalized in the estimation process. By doing so, Problem (\ref{eq:11}) becomes
\begin{equation} \label{eq:13}
\begin{array}{ll}
\textrm{MLE:}\quad\underset{\bR \in Toep, \bR \succ 0}{\rm minimize} \: {\bar{f}}(\bR),
\end{array}
\end{equation}
where $Toep$ is used to denote the set of {Hermitian} Toeplitz matrices of size $m \times m$. The above problem has two constraints: a structural constraint and a positive definite constraint. Even though the structural constraint is convex, the non-convexity of the objective function makes Problem (\ref{eq:13}) challenging to solve and no analytical solution seems to be available. In the following two iterative solution procedures for (\ref{eq:13}) are designed exploiting the MM principle. Briefly, the MM technique mainly consists of two steps
\begin{enumerate}
	\item constructing a s.f. $g(\bR|\bR_{t})$ (where $\bR_{t}$ is the estimate of $\bR$ at the $t^{th}$ iteration) for the objective function in (\ref{eq:13});
	\item minimizing the resulting surrogate problem at each iteration.
\end{enumerate}
For more details, \cite{MM1, conv, am} provide an in-depth discussion on MM based algorithms.

\section{Algorithms for Toeplitz covariance matrix estimation}\label{sec:3}
In this section, ATOM1 and ATOM2 are proposed to tackle the MLE problem of TSC matrix. Both {exploit} the MM principle (applied to an equivalent reformulation of the MLE problem) and differ in the way they construct and handle the surrogate minimization problem. ATOM1 solves the surrogate optimization using ADMM while ATOM2 tackles it using {either} alternating projection or Dykstra's  algorithm. {Subsequently,} the computational complexity and proof of convergence of the procedures are established. Finally, the extension of ATOM2 to deal with additional covariance constraints along with the  Toeplitz structure is provided.

{Before proceeding further, let us observe that the Hermitian Toeplitz matrices intrinsically endow the centro-Hermitian symmetry structure~\cite{1093391}, i.e.,
	\begin{equation}
{\bR = \bJ \bR^* \bJ}
	\end{equation}
{with $\bJ$ the $m\times m$ permutation matrix given by
\begin{equation}\label{eq:J_exchange_matrix}
	\bJ = \begin{bmatrix}
		0 & 0 & \cdots & 0 & 1 \\ 
		0 & 0 & \cdots & 1 & 0\\ 
		\vdots & \vdots & \ddots & \vdots & \vdots\\ 
		1 & 0 & \cdots & 0 & 0
	\end{bmatrix} .
\end{equation}}
As a consequence, Problem~\eqref{eq:13} is tantamount to 
\begin{equation}\label{eq:minimize_prob}
	\underset{\bR \in Toep, \bR \succ 0}{\rm minimize} \: f(\bR),
\end{equation}
where
\begin{equation}\label{eq:obj_function}
	f(\bR) = \trace(\bR_{FB} \bR^{-1}) + \log|\bR|
\end{equation}
refers to the restriction of $\bar{f}(\cdot)$ to the centro-Hermitian covariance matrices, with $\bR_{FB}$ the forward-backward (FB) averaged sample covariance matrix\footnote{{Hereafter, Problem~\eqref{eq:13} (and thus~\eqref{eq:minimize_prob}) is assumed solvable, i.e., there exists a global optmizer $\bR^* \succ 0$, as well as any limit point of a feasible sequence of matrices whose corresponding objectives converge to the optimal value is feasible to the optimization problem. As a consequence, without loss of generality, the constraint $\bR \succ 0$ can be relaxed into $\bR \succeq 0$. Notably, a sufficient condition to ensure the aforementioned properties is provided by $n \ge \lceil m/2 \rceil$, corresponding to $\bR_{FB} \succ 0$ with probability one.}} given by $\bR_{FB} = 1/2 (\bR_{\textrm{SCM}} + \bJ \bR_{\textrm{SCM}}^* \bJ)$~\cite{vantrees4}.}

{Now, decomposing $\bR_{FB}=\bY \bY^H$, e.g., via LDL factorization, with $\bY \in \mathbb{C}^{m \times r}$, where $r=rank(\bR_{FB})\le m$, Problem~\eqref{eq:minimize_prob} can be equivalently cast as\footnote{{A similar constraint reformulation is used in some studies involving atomic norm for sparse reconstruction~\cite{6560426, 7314978}.}} (the interested reader may refer to Appendix A of the supplementary material to this paper)}
\begin{equation}
\begin{aligned}\label{Opt:1}
\min_{\bR \in Toep,\bX{\in \mathbb{H}^{r\times r}}}& \trace(\bX) + \log|\bR|\\
\text{s.t.}~~~~& \left(\begin{array}{c c}
\bX&\bY^H\\
\bY&\bR
\end{array}\right)\succeq\mathbf{0}
\end{aligned},
\end{equation}
{where the objective is a concave differentiable function of $\bX$ and $\bR$.}

Before proceeding with the next important lemma, it is worth pointing out that Problem \eqref{Opt:1} holds true even if the Toeplitz structural constraint in Problem \eqref{eq:13} and \eqref{Opt:1} is replaced by {any set of positive definite matrices, provided that the estimation problem is solvable}.

\begin{lemma}\label{lemma 0}
Given a concave differentiable\footnote{For a non-differentiable function, the inequality in (\ref{l0}) can be {cast as} ${h}(\bX) \leq {h}(\bX_{t}) + {\rm{Tr}}\left(\bG(\bX_{t})^{H} (\bX-\bX_{t})\right)$, where $\bG(\bX_{t})$ is the subgradient of the concave function ${h}(\bX)$ at $\bX_{t}$~{\cite{MM1}}. } function ${h}(\bX): \mathbb{H}^{r \times r} \rightarrow \mathbb{R}$, it can be majorized as
\begin{equation}\label{l0}
    \begin{array}{ll}
    {h}(\bX) \leq {h}(\bX_{t}) + {\rm{Tr}}\left(\nabla {h}(\bX_{t})^{H} (\bX-\bX_{t})\right),
    \end{array}
    \end{equation}
where ${\bX}_{t} {\in \mathbb{H}^{r \times r} }$. The upper bound to ${h}(\bX)$ is linear and differentiable with respect to (w.r.t.) $\bX$.
\begin{IEEEproof}
Since ${h}(\bX)$ is a concave function w.r.t. $\bX$, (\ref{l0}) stems from linearizing ${h}(\bX)$ via its first order Taylor expansion~\cite{matrix_taylor}.
\end{IEEEproof}
\end{lemma}

{In order to tackle the challenging optimization problem~\eqref{Opt:1}, MM-based methods~\cite{Wu-MM, heiser1995convergent}, {denoted ATOM1 and ATOM2, are} now developed.}
{To this end, let us observe that the term} $\log|\bR|$ in \eqref{Opt:1} is a concave function w.r.t. $\bR$ \cite{boyd}. {Hence, it can be majorized using Lemma~\ref{lemma 0} to get the following s.f.}
\begin{equation}\label{Opt:2}
\begin{aligned}
		g(\bX,\bR|\bR_t) &=\trace(\bX) + \trace\big(\bR_t^{-1}\bR\big) + c_1 \\
						 &={ \trace(\bA_t\bE) + c_1}
\end{aligned},
\end{equation}
{where the constant $c_1 = \log|\bR_t| - m$, ${\bA}_{t} = \text{diag}(\bI,\bR_{t}^{-1})$, whereas $\bE = \text{diag}(\bX,\bR)$ is the block-diagonal matrix with blocks $\bX$ and $\bR$ along the main diagonal}. Given $\bR_{t}$, {which in our case is the value assumed by the variable $\bR_{t}$ at the $t$-th iteration of the algorithm,} the MM method demands for the solution of  the following surrogate minimization {task}
\begin{equation} \label{s1}
\begin{aligned}
\bR_{t+1} \quad = \quad &\underset{\bR\in Toep, \bX {\in \mathbb{H}^{r\times r}}}{\rm arg\: min} g(\bX,\bR|\bR_t) \\ & {\rm s.t.} \:\left(\begin{array}{l l}
	\bX&\bY^H\\
	\bY&\bR \end{array}\right) \succeq \bzero
\end{aligned},
\end{equation}
{which} is a SDP problem. {Unfortunately}, the computational complexity necessary to handle SDP using interior point methods is $\mathcal{O}(({r}+{m})^{{6.5}})$ \cite{sdpcomplexity}. {In order to alleviate the computational issue, two different approaches are pursued. The former directly handles Problem~\eqref{s1} via the iterative ADMM algorithm. The latter, by means of a suitable manipulation of~\eqref{Opt:2}, constructs a different s.f. for the objective function in Problem~\eqref{Opt:1}. By doing so, as clearly explained in the following, a computationally {efficient} and flexible estimation procedure capable of including additional constraints can be developed. To this end, let us observe that, adding and subtracting $\gamma \textrm{Tr}({\bE}^{2})$,~\eqref{Opt:2} {is equivalent to} 
\begin{equation}\label{eq_trace1}
		\trace(\bA_t\bE) + \gamma\textrm{Tr}({\bE}^{2})-\gamma\textrm{Tr}({\bE}^{2}) 
\end{equation}
with $\gamma > 0\in \mathbb{R}$ a parameter of the surrogate construction stage (for $\gamma \downarrow 0$, the function in~\eqref{eq_trace1} reduces to~\eqref{Opt:2}).
}
{
Now, being $-\textrm{Tr}({\bE}^{2})$ a concave function of $\bE$ and invoking Lemma~\ref{lemma 0} applied to the feasible solution $\bE_{t}=\text{diag}(\bX_t;\bR_t)$ with $\bX_t = \bY^H\bR_t^{-1}\bY$ and $\bR_t$ provided by the $t$-th iteration step of the estimation process, it is possible to construct the s.f. for~\eqref{eq_trace1}
\begin{equation}\label{ns}
	\begin{aligned}
		\tilde{g}(\bX,\bR|\bR_{t}) = & \textrm{Tr}\left({\bA_t}{\bE}\right)+\gamma \textrm{Tr}({\bE}^{2})-2\gamma\textrm{Tr}({\bE}{{\bE}}_{t})\\& - \gamma \textrm{Tr}(\bE_t^2).
	\end{aligned}
\end{equation}
}

{
It is worth pointing out that $	\tilde{g}(\bX,\bR|\bR_{t})$ represents a surrogate to a s.f.. Nonetheless, since $	\tilde{g}(\bX,\bR|\bR_{t})$ is a tight approximation of $g(\bX,\bR|\bR_{t})$, it is straightforward to show that~\eqref{ns} provides a direct surrogate for the objective function in Problem~\eqref{Opt:2}. Hence, given $\bR_{t}$ and after some algebraic manipulations, the resulting surrogate minimization problem at the $t$-th iteration can be cast as 
\begin{equation} \label{prefinal}
	\begin{aligned}
			\bR_{t+1}=&\underset{{\bR}  \in Toep,\bX}{\rm arg\:min} \: \|\bE - \bB_t\|_{F}^{2}\\
		&{\rm subject\ to}\quad \bE+\bD \succeq \mathbf{0}
	\end{aligned},
\end{equation}
where $\bB_t = {{\bE}}_{t} - \gamma'{\bA}_{t}$, with $\gamma' = \frac{0.5}{\gamma}$ and $\bD=[\bzero,\bY^H;\bY,\bzero]$.
}

{In the following subsections~\ref{sec:3a} and \ref{sec:3c} two iterative methods{, i.e., ATOM1 and ATOM2,} are proposed to solve the surrogate minimization problems in~\eqref{s1} and~\eqref{prefinal}, respectively.}
\subsection{ATOM1}\label{sec:3a}
The surrogate minimization problem in (\ref{s1}) is solved using ADMM \cite{admmmatrix1,admmmatrix2}. To {this end}, an auxiliary variable $\bU {\in \mathbb{H}^{r+m \times r+m}}$ is introduced in (\ref{s1}) and the problem is {framed} in the equivalent form
\begin{equation}
\begin{aligned}\label{admm1}
\min_{\bR \in Toep,\bU \succeq \bzero,\bX{\in \mathbb{H}^{r\times r}}}& \trace(\bX) + \trace\big((\bR_t)^{-1}\bR\big)\\
\text{s.t.}~~&~~ \left(\begin{array}{c c}
\bX&\bY^H\\
\bY&\bR
\end{array}\right)-\bU =\mathbf{0}
\end{aligned}.
\end{equation}
The augmented Lagrangian {associated with} (\ref{admm1}) is
\begin{align}\label{Lagrange:1}
&\mathcal{L}_{\rho}(\bR,\bX,\bU,\hat{\blam})=\trace(\bX) + \trace\big((\bR_t)^{-1}\bR\big) \nonumber\\
&\quad + \trace\left[\hat{\blam}^H\left(\left(\begin{array}{c c}
\bX&\bY^H\\
\bY&\bR
\end{array}\right)-\bU\right)\right]\nonumber\\
&\quad+ \frac{\rho}{2}\left\Vert\left(\begin{array}{c c}
\bX&\bY^H\\
\bY&\bR
\end{array}\right)-\bU\right\Vert_F^2,
\end{align}
where $\rho >0$ is the penalty parameter and $\hat{\blam}$ is the Lagrange multiplier of size {$(r+m)\times (r+m)$}. Problem \eqref{Lagrange:1} can be further rewritten as
\begin{align}
 \mathcal{L}_{\rho}(\bE, \bU, \hat{\blam}) &= \trace(\bA_t \bE) + \trace\big(\hat{\blam}^H (\bE + \bD - \bU)\big)\nonumber\\
 &\quad + \frac{\rho}{2}\Vert\bE + \bD - \bU\Vert_F^2.
\end{align}
The (inner) iterative steps of ADMM algorithm \cite{admmmatrix1,admmmatrix2} are
\begin{align}
\bU_{k+1}^t &= \underset{\bU \succeq \bzero}{\arg\min}\quad \trace\big((\hat{\blam}_k^t)^T (\bE_k^t + \bD - \bU)\big)\nonumber\\
&\quad + \frac{\rho}{2}\Vert\bE_k^t + \bD - \bU\Vert_F^2\label{updateU}
\end{align}
\begin{align}
\bE_{k+1}^t& = \underset{\bR\in Toep,\bX}{\arg\min}\quad \trace(\bA \bE) + \trace\big((\hat{\blam}_k^t)^T (\bE + \bD - \bU_{k+1}^t)\big)\nonumber\\
&\quad + \frac{\rho}{2}\Vert\bE + \bD - \bU_{k+1}^t\Vert_F^2\label{updateE}\\
\hat{\blam}_{k+1}^t&=\hat{\blam}_{k}^t + \rho \big(\bE_{k+1}^t + \bD - \bU_{k+1}^t\big)\label{updateL},\quad\quad\quad\quad\quad\quad
\end{align}
where $(\cdot)^{t}_{k}$ is used to denote the $k$-th {inner-}iteration of the ADMM algorithm in correspondence of the $t$-th MM outer-loop. Problems (\ref{updateU}) and (\ref{updateE}) have closed-form solutions which can be computed via the projection of appropriate matrices onto the respective feasible sets. Indeed, Problem (\ref{updateU}) can be equivalently cast as
\begin{equation}\label{newupdateX}
\begin{array}{ll}
\bU^{t}_{k+1}= \underset{{\bU \succeq 0 }}{\rm arg\:min}\: \|\bU -   \bPsi^{{t}}_{{k}}\|_{F}^{2}\\
\end{array}
\end{equation}
where $\bPsi^{{t}}_{{k}} = \bE_k^t + \bD + \frac{1}{\rho}\hat{\blam}_k^t$. Hence, solving (\ref{updateU}) is tantamount to performing the orthogonal projection of the matrix  $\bPsi^{{t}}_{{k}}$ onto the set of the {PSD} matrices which can be computed as $\bU^{t}_{k+1}=\tilde{\bV}^{{t}}_{{k}}\max({\operatorname{diag}}(\tilde{\bU}^{{t}}_{{k}}),\bzero)\tilde{\bV}^{{t H}}_{{k}}$, where ${\operatorname{diag}}(\tilde{\bU}^{{t}}_{{k}})$ and $\tilde{\bV}^{{t}}_{{k}}$ are the matrices containing the eigenvalues and the corresponding {orthonormal} eigenvectors of $\bPsi^{{t}}_{{k}}$, respectively. Similarly, the update step of $\bE$ in (\ref{updateE}) can be rewritten as
\begin{align}\label{newupdateR}
\bE_{k+1}^t = \underset{\bR\in Toep,\bX}{\arg\min}~~ \Vert\bE  - \bLam{_{k}^t} \Vert_F^2,
\end{align}
where {$\bLam{_{k}^t} =  {\mathcal{P}_{{D–}Toep}}\left( \bU_{k+1}^t-\bD - \frac{1}{\rho} (\hat{\blam}_k^t +\bA{_{t}})\right)$, with ${\mathcal{P}_{{D–}Toep}}(\bPsi)$ computed as follows:} {Partitioning the matrix $\bPsi$ as} $\bPsi=\left(\begin{array}{c c}
		\bPsi_{11}&\bPsi_{12}\\
		{\bPsi^H_{12}}&\bPsi_{22}
	\end{array}\right)$ with $\bPsi_{12}$ of size ${r}\times m$, the orthogonal projection {of interest amounts to set the upper diagonal block to $\bPsi_{11}$ whereas} the second diagonal block is obtained by averaging the elements along each diagonal of $\bPsi_{22}$ {and constructing the corresponding Toeplitz matrix}.

 {Now, partitioning $\bLam{_{k}^t}$ as} $\bLam{_{k}^t} = \left(\begin{array}{c c}
\bLam^{{t}}_{11{,k}}&\bLam^{{t}}_{12{,k}}\\
\bLam^{{tH}}_{12{,k}}&\bLam^{{t}}_{22{,k}}
\end{array}\right)$
with $\bLam^{{t}}_{11{,k}}$ and $\bLam^{{t}}_{22{,k}}$ being {$r \times r$} and $m \times m$ matrices, respectively, it follows that {$\bX_{k+1}^t=\bLam^{{t}}_{11{,k}}$ and $\bR_{k+1}^t = \bLam^{{t}}_{22{,k}}$.}
Before concluding, it is worth pointing out that since the surrogate minimization problem in (\ref{s1}) is convex {and only an equality constraint is forced}, it is guaranteed that ADMM converges to a supposed existing\footnote{{A sufficient condition for the existence of the optimal solution to Problem~\eqref{s1} is provided by the solvability of~\eqref{eq:minimize_prob}}.} optimal unique solution to (\ref{s1}) (see Section $3.2$ in \cite{boydadmm}, \cite{admmnew}). The pseudocode of the proposed algorithm is shown in {\textbf{Algorithm 1}}.\\
From Algorithm 1 it can be seen that ATOM1 requires initialization of the matrices $\bR_{0}$, $\bX^{t}_{0}$ and $\hat{\blam}^{t}_{0}$. $\bR_{0}$ can be set using the initialization scheme discussed in \cite{melt} and, as $t=0$, $\bX^{t}_{0}$ can be set equal to $\bY^H\bR_0^{-1}\bY$ while $\hat{\blam}^{t}_{0}$ can be constructed as $\hat{\blam}^{t}_{0} ={ \bV\bV^{H}}$, where the elements of $\bV$ are drawn randomly from a uniform distribution over $[0,1]$. For $t\geq1$, the matrices $\bE^{t}_{0}$ and $\hat{\blam}^{t}_{0}$ can be initialized with their last value after convergence at the previous ADMM iteration, respectively. Another input parameter required by ATOM1 is the penalty {weight} $\rho$, introduced during the {construction of the} Augmented Lagrangian of  the ADMM {framework}. It is shown in \cite{boydadmm}, that the ADMM algorithm converges for any value of $\rho>0$. However, the numerical stability and the convergence rate depends on the choice of $\rho$. Simulation results have highlighted that for $\rho = 1$, the ADMM algorithm is stable for different values of $n$ and $m$. Hence, unless otherwise stated, in all the numerical analysis $\rho = 1$ is used.
\begin{table}[t]
\begin{center}
	\resizebox{1\linewidth}{!}{
{\begin{tabular}{l}
	\hline
{\bf{Algorithm 1}} Pseudocode of ATOM1 algorithm \\
\hline
{\bf{Input}}: Data-based matrix  $\bY$ and $\rho$ \\
{\bf{Initialize}}: Set ${t}, {k} = 0$. Initialize ${{\bR}_{0}}$, ${{\bX}_{0}}$ and $\hat{\blam}_{0}$.\\
{\bf{Repeat}}:\\
\hspace{5mm}$k \leftarrow 0$\\
\hspace{4mm} Compute $\bA_t = \text{diag}(\bI,\bR_{t}^{-1})$, ${{\bE}}^{t}_{k} = \text{diag}(\bX_t,\bR_t)$, $\hat{{\blam}}^{t}_{k} = \hat{\blam}_{t}$\\
\hspace{5mm}{\bf{Repeat}}:\\
\hspace{9mm}1) Obtain $\bU^{t}_{k+1}$ by projecting the matrix \\ \hspace{11mm} $\bPsi^{{t}}_{{k}} = \bE_k^t + \bD + \frac{1}{\rho}\hat{\blam}_k^t$ onto the set of PSD matrices.\\
\hspace{9mm}2) Compute $\bLam = \bU_{k+1}^t-\bD - \frac{1}{\rho} (\hat{\blam}_k^t +\bA_t)$\\
\hspace{9mm}3) Set $\bX^{t}_{k+1}$ equal to the first block $\bLam_{11}$ of $\bLam$ \\
\hspace{9mm}4) Obtain $\bR^{t}_{k+1}$ by projecting the second block $\bLam_{22}$ of $\bLam$ \\
\hspace{11mm} onto the set of Toeplitz matrices.\\
\hspace{9mm}5) Obtain ${{\bE}}^{t}_{k+1} = \text{diag}(\bX_{k+1}^t,\bR_{k+1}^t)$\\
\hspace{9mm}6) $\hat{\blam}^{t}_{k+1} = \hat{\blam}^{t}_{k} +\rho\big(\bE_{k+1}^t + \bD - \bU_{k+1}^t\big)$\\
\hspace{9mm}{7) $k\leftarrow k+1$}\\
\hspace{4mm} {\textbf{until convergence}} \\
\hspace{4mm} Set {${\bR}_{t+1}= {\bR}^{t}_{k}$}, {${\bX}_{t+1}= {\bX}^{t}_{k}$}, $\hat{\blam}_{t+1}= \hat{\blam}^{t}_{k}$ \\
\hspace{5mm}$t\leftarrow t+1$\\
{\textbf{until convergence}}\\
{\bf{Output}}: $\bR_{{\rm{\bf{ATOM1}}}}= \bR_{t}$.\\
\hline
\end{tabular}}}
\end{center}
\end{table}
\subsubsection{\textbf{Computational complexity and discussion about ATOM1}}\label{sec:3b}
ATOM1 is iterative in nature with two loops - the outer-loop updates the Toeplitz matrix $\bR_{t}$ while the inner-loop solves the surrogate minimization problem using ADMM. Note that in the inner-loop, {it is required to} construct the data-based matrix $\bD = \left(\begin{array}{c c}
\mathbf{0}&\bY^H\\
\bY&\mathbf{0}
\end{array}\right)$ - which is iteration independent and hence can be pre-{computed and stored}.
Let us now discuss the complexity related to the outer and inner-loops of ATOM1. The inner-loop of ATOM1 requires the computation of the matrix {$\bA_t$} - which is outer-loop iteration dependent. Therefore, this matrix can be {evaluated} once in {each} outer-loop. Consequently, apart from the computations {involved in} the inner-loop, an outer-loop cycle {just involves the evaluation} of the matrix $\bR_{t}^{-1}$. {Since $\bR_{t}$ is Toeplitz, its inverse can be efficiently computed with a complexity $\mathcal{O}(m\:{\rm{log}}m)$ \cite{fast-inversion-Toeplitz}.} The computational complexity of an inner-loop cycle is related to the projection of $\bPsi_{{k}}^{{t}}$ onto the set of PSD matrices and projection of $\bLam^{{t}}_{{k}}$ onto {the set of block diagonal matrices where the upper part (of size $r \times r$) is unconstrained, whereas the lower block (of size $m \times m$) is Toeplitz structured.}
The cost of {this latter operation mainly involves the projection of} $\bLam^{{t}}_{22{,k}}$ onto the set of Toeplitz matrices{; thus, it is substantially} dictated by the computation of average of the elements along the diagonals of $\bLam^{{t}}_{22{,k}}$. Hence, the cost {of the inner-step \textit{4)} is} $\mathcal{O}(m^{2})$. Next, the projection of $\bPsi$  onto the set of PSD matrices mainly involves the computation of the eigenvalues and eigenvectors of the matrix $\bPsi_{{k}}^{{t}}$ - whose corresponding complexity is $\mathcal{O}(({r+m})^{3})$ \cite{golub}. Therefore, the {per-outer-iteration} computational complexity of ATOM1 is $\mathcal{O}(\eta({r+m})^{3})$ where $\eta$ is the total number of inner-loop iterations required by the algorithm to converge.

A drawback of ATOM1 is {the lack of a theoretical quality guarantee when it has to handle} additional constraints on the covariance matrix. This is because ATOM1 implements ADMM algorithm at {each} inner-iteration which requires {(to endow convergence guarantees to the process)} the optimization problem to exhibit the standard form \cite{boydadmm,admmsdp}
\begin{equation}\label{format}
\begin{array}{ll}
\underset{{\bZ}, \bE}{\rm minimize} \:& {h_1}({\bZ_1}) + {h_2}({\bZ_2}) \\
{\rm subject\ to} & \:\: \bA_{1}{\bZ_1}+\bA_{2}{\bZ_2} = \bC
\end{array}
\end{equation}
where ${h_1}({\bZ_1})$, ${h_2}({\bZ_2})$ are convex functions and $\bA_{1}$, $\bA_{2}$, $\bC$ are matrices of appropriate dimensions, respectively. Therefore, to incorporate additional inequality constraints (such as {those resulting from} upper bound on the condition number of the matrix ${\bZ_1}$ {or a lower bound to the strength of diagonal elements, or more in general an intersection of closed convex sets that can be described by additional auxiliary variables}), one needs to replace {each} inequality constraint with {an appropriate equality constraint}. This can be done by introducing a slack variable {for each inequality constraint to} the existing optimization variables ${\bZ_1}$ and ${\bZ_2}$. However, there is no convergence guarantee of ADMM when there are more than two optimization variables \cite{admmext}. This issue {can be addressed} by {the} {low complexity} algorithm, referred to as ATOM2, {proposed to} solve {Problem~\eqref{prefinal}}.
\subsection{ATOM2}\label{sec:3c}
{Problem~\eqref{prefinal} is tantamount to seeking the block diagonal matrix $\bE$ belonging to the intersection of the two sets - the former defined by block diagonal matrices with the lower diagonal block of size $m \times m$ fulfilling a Toeplitz structure and the latter given by the Linear Matrix Inequality (LMI)~\cite{LMI_book} $\bE + \bD \succeq 0$ -  with minimum distance from $\bB$. Being the feasible set of~\eqref{prefinal} characterized by the intersection of convex sets, a viable, even though heuristic, means to tackle Problem~\eqref{prefinal} is provided by the alternating projection or Projection Onto the Convex Sets (POCS) technique~\cite{ap1, pocsnew, pocs}, which has already been successfully applied  in the signal processing context, e.g.,~\cite{1377501, 1677924}.
}

Let us denote by ${\mathcal{P}_{LMI}}(\bPsi)$ the orthogonal projection of an arbitrary matrix $\bPsi$ onto the set defined by $\bE+\bD \succeq \mathbf{0}$. Now, to proceed further and employ the POCS framework, ${\mathcal{P}_{D–Toep}}(\bPsi)$ and ${\mathcal{P}_{LMI}}(\bPsi)$ projections must be employed. Remarkably, both can be obtained {in closed-form: the former is computed as described in subsection~\ref{sec:3a}; as to the latter, the} orthogonal projection onto the set defined by LMI $\bE+\bD \succeq \mathbf{0}$ {is computed by first evaluating} the EigenValue Decomposition (EVD) of the matrix $\bPsi +\bD$, i.e., {obtaining} $[\bar{\bU}, \bar{\bV}] = {\rm{eig}}(\bPsi +\bD)$, where $\bar{\bU}$ and $\bar{\bV}$ are matrices containing the eigenvalues and eigenvectors  of the spectral decomposition, respectively. Then, the orthogonal projection ${\mathcal{P}_{LMI}}(\bPsi)$ is given by $\bar{\bV}{\rm{max}}(\bar{\bU},\bzero)\bar{\bV}^{H} - \bD$.

According to POCS method, given an initial value $\bT_{0}^{{t}} = \bB_{{t}}$, {at the $k$-th inner-iteration} first compute $\bY^{{t}}_{k+1} ={\mathcal{P}_{{D–}Toep}}(\bT^{{t}}_{k})$ and then, using $\bY^{{t}}_{k+1}$, determine $\bT^{{t}}_{k+1}={\mathcal{P}_{LMI}}(\bY^{{t}}_{k+1})$ {which represents} the starting point {$\bT^{{t}}_{k+1}$} of the next {inner-}iteration. Hence, {the} POCS{-based solution approach} finds a sequence of iterates $\{\bT^{{t}}_{k}\}$ by alternatingly projecting between the two convex sets.  {Nevertheless}, as reported in \cite{slowconvg}, POCS may suffer from slow convergence. {Even more crucial, the convergence to the global optimal solution to~\eqref{prefinal} is, in general, not ensured{~\cite{dykstra, dykstra2}.}} {A possible solution to the aforementioned shortcoming is provided by Dykstra's} projection \cite{dykstra} {which is a refinement of POCS capable of finding} a point closest to $\bB_{{t}}$ by adding correction {matrices} $\bP_{k}$  and $\bQ_{k}$ before {each} projection {is performed}, which in-turn ensures convergence of sequence $\{\bT_{k+1}\}$ to the optimal solution ${\bT}^{*}{=\bE^{*}}$ \cite{dykstra}. The pseudocode of Dykstra's algorithm is shown in {\textbf{Algorithm 2}}.
\begin{table}[t]
\begin{center}
\resizebox{0.9\linewidth}{!}{
\begin{tabular}{l}
\hline
{\bf{Algorithm 2}} Pseudocode of Dykstra's  algorithm \\
\hline
{\bf{Input}}: $\bB_{{t}}$ \\
{\bf{Initialize}}: Set $\bT_{0}^{{t}}= \bB_t$, ${{\bP}^{{t}}_{0}}=\bzero$ and ${{\bQ}^{{t}}_{0}}=\bzero$, $k=0$\\
{\bf{Repeat}}: \\
\hspace{5mm}1) $\bY_{k}^{{t}} = {\mathcal{P}_{{D–}Toep}}(\bT^{{t}}_{k}+\bP^{{t}}_{k})$\\
\hspace{5mm}2) $\bP^{{t}}_{k+1} = \bT^{{t}}_{k}+\bP^{{t}}_{k}-\bY^{{t}}_{k}$\\
\hspace{5mm}3) $\bT^{{t}}_{k+1} = {\mathcal{P}_{LMI}}(\bY^{{t}}_{k}+\bQ^{{t}}_{k})$\\
\hspace{5mm}4) $\bQ^{{t}}_{k+1} = \bY^{{t}}_{k}+\bQ^{{t}}_{k}-\bT^{{t}}_{k+1}$\\
\hspace{5mm}5) $k\leftarrow k+1$\\
{\bf{until convergence}}\\
{\bf{Output}}: ${\bE}^{*} =\bT^{{t}}_{k}$.\\
\hline
\end{tabular}}
\end{center}
\end{table}
Once the optimal solution $\bE^{*}$ is obtained via Dykstra's projection, the matrix ${\bR}_{t+1}$ can be constructed from its lower {diagonal} block of size $m \times m$. This process is repeated until the whole {MM-procedure}, i.e., including the outer-loop, converges.
{The complete} ATOM2 is summarized in {\textbf{Algorithm 3}}.
\begin{table}[t]
\begin{center}
\resizebox{0.96\linewidth}{!}{
\begin{tabular}{l}
\hline
{\bf{Algorithm 3}}  Pseudocode of ATOM2 \\
\hline
{\bf{Input}}: {Data-based matrix  $\bY$, surrogate parameter $\gamma$} \\
{\bf{Initialize}}: Set \emph{t} = 0. Initialize $\bR_{0}$, ${{\bX}_{0}}$. \\
{\bf{Repeat}}:\\
\hspace{5mm}1) Compute $\bA_t = \text{diag}(\bI,\bR_{t}^{-1})$, ${{\bE}}_t = \text{diag}(\bX_t,\bR_t)$\\
\hspace{5mm}2) Compute $\bE^{*}$ from {{\textbf{Algorithm 2}}} execution with\\
\hspace{8mm}$\bB_{{t}} = {\bE}_{t} - \frac{0.5}{\gamma}{\bA_{{t}}}$\\
\hspace{5mm}3) Obtain $\bR_{t+1}$ from the lower {diagonal} block of $\bE^{*}$ \\\hspace{5mm}4) Obtain $\bX_{t+1}$ from the upper {diagonal} block of $\bE^{*}$ \\
\hspace{5mm}5) $t\leftarrow t+1$\\
{\bf{until convergence}}\\
{\bf{Output}}: $\bR_{{\rm{\bf{ATOM2}}}}=\bR_t$\\
\hline
\end{tabular}}
\end{center}
\end{table}
{It} requires the initialization of the matrix $\bR$. In this respect, a similar scheme as in ATOM1 is followed, i.e., at each outer-iteration, the initial guess required to determine $\bR_{t+1}$ in the inner-loop is {obtained starting from} $\bR_{t}$.
\subsection{Computational complexity of ATOM2}\label{sec:3d}
{Like} ATOM1, ATOM2 is an iterative algorithm with outer- and inner-loops. The outer-loop updates the Toeplitz matrix $\bR_{t}$ and the inner-loop implements the Dykstra's algorithm - which requires the computation of the matrices $\bD$ and ${\bR_{t}^{-1}}$. The former is a iteration independent data matrix and therefore can be pre-constructed. {The latter is} outer-loop iteration dependent and therefore can be computed once in {each} outer-loop. Consequently, apart from the inner-loop computations, the outer-loop demands only the
computation of ${\bR_{t}^{-1}}$ -  which can be computed efficiently with complexity $\mathcal{O}(m\:{\rm{log}}m)$. Meanwhile, the computational load of the inner-loop stems from the {evaluation} of EVD of the matrix $(\bY_{k} +\bQ_{k})$ plus a data matrix $\bD$ - which has a complexity of about $\mathcal{O}(({r+m})^{3})$.

In Table \ref{c1}, the computational complexity of ATOM1 and ATOM2 is compared with that of the state-of-the-art iterative algorithms \cite{melt,em1}. Unlike the proposed algorithms, the state-of-the art methods are single loop iteration algorithms. Therefore, in the case of \cite{melt,em1} $\eta$ is used to represent the number of iterations required by the algorithm to converge. Inspection of Table \ref{c1} shows that ATOM1 and ATOM2 {have} the highest complexity when compared to MELT and EM. Nevertheless, it is worth anticipating that this complexity increase is complemented by a superior performance in terms of {generality of the problem solved (ATOM1 and ATOM2 do not exploit the CE, ATOM2 permits to handle additional structural constraints with quality guarantee, as shown in subsection~\ref{sec:3f})}, covariance matrix MSE, and achieved SINR.

\begin{table}[t]
	\centering
	\caption{Comparison among computational complexity of ATOM1 and ATOM2 with other state-of-the-art iterative algorithms.}
	\resizebox{0.6\linewidth}{!}{
	\begin{tabular}{ll}
		\hline\hline
		\textbf{Algorithm} & \textbf{Complexity} \\ \hline
		ATOM1& $\mathcal{O}(\eta({r+m})^3))$\\
		 ATOM2& $\mathcal{O}(\eta({r+m})^3))$ \\
		 MELT \cite{melt}& $\mathcal{O}\left(\eta(m{\rm{log}}(m))\right)$ \\
		 EM \cite{em1} & $\mathcal{O}\left(\eta(m{\rm{log}}(m)\right))$\\
		\hline		\hline
	\end{tabular}}
	\label{c1}
\end{table}
\subsection{Proof of convergence}\label{sec:3e}
In this subsection, the proof of convergence of ATOM1 and ATOM2 is established. In this regard, it is worth pointing out that both the algorithms differ in the way they construct and optimize the s.f. for the Problem \eqref{eq:13}. Nonetheless, since ATOM1 and ATOM2 are based on the MM framework, the proof of convergence based on the following {Theorem} will hold for both algorithms. \\
Before stating the {Theorem}, let us first introduce the first-order optimality condition for minimizing a function over a convex constraint set.  A point $\bX$ is a stationary point of $f(\cdot)$ if $f'(\bX;\bD) \geq 0$ for all $\bD$ such that $\bX+\bD \in \mathcal{C}$, where $\mathcal{C}$ is the convex constraint set and  $f'(\bX;\bD)$ is the directional derivative of $f(\cdot)$ at point $\bX$ in direction $\bD$ and is defined as \cite{conv}
\begin{equation}
\begin{array}{ll}
f'(\bX;\bD) =\underset{\lambda \downarrow 0}{\lim} \: \textrm{inf}\:\dfrac{f(\bX+\lambda\bD) - f(\bX)}{\lambda}
\end{array}.
\end{equation}
Based on the following {Theorem}, both ATOM1 and ATOM2 are guaranteed to converge to a stationary point of Problem~{\eqref{prefinal}}.
\begin{theorem}\label{lemma 3}
Denoting by $\{\bR_{t}\}$ the sequence {of matrices} generated by either ATOM1 or ATOM2, then the objective function of Problem \eqref{eq:13} monotonically decreases along the iterations. Besides, any positive definite cluster point\footnote{{Under the assumption $m\ge n/2$, all the cluster points are demanded to be positive definite.}} to $\bR_{t}$ is a stationary point to Problem \eqref{eq:13}.
\begin{IEEEproof}
See Appendix B of the supplementary material for details.
\end{IEEEproof}
\end{theorem}

\subsection{{Extensions of ATOM2}}\label{sec:3f}
The {augmentation} of ATOM2 to handle additional constraints other than the Toeplitz structure in the covariance estimation process is now addressed. In particular, it is shown that ATOM2 can be generalized to account for the following scenarios: {Banded Toeplitz, block-Toeplitz, and Toeplitz-block-Toeplitz matrices}. {On the other side, as already mentioned in subsection~\ref{sec:3b}},  ATOM1 cannot be directly extended to tackle the general constraints as for instance an upper bound requirement to the condition number.

\subsubsection{MLE of banded Toeplitz covariance matrix}\label{subsection:ATOM2_bandedT}
The covariance matrix is constrained to exhibit a banded Toeplitz structure of bandwidth $b$ (see \cite{band, bt} for relevant applications). For instance, assuming a bandwidth $b=2$ and dimension $m=5$ the covariance matrix enjoys the following structure
\[ \bR=
\begin{bmatrix}
    r_{1} & r_{2} & r_{3}& 0 & 0 \\
    r^{*}_{2} & r_{1}  & r_{2}  & r_{3} & 0\\
    r^{*}_{3} &  r^{*}_{2}  & r_{1}  & r_{2} &  r_{3}\\
    0 &  r^{*}_{3}  & r^{*}_{2}    & r_{1} &  r_{2}\\
    0 &  0& r^{*}_{3}  & r^{*}_{2}    & r_{1}
\end{bmatrix}.
\]
Then, the MLE problem for banded Toeplitz covariance matrix can be formulated as
\begin{equation}
\begin{array}{ll}
 \underset{\bR  \in Band-Toep,\:\bR \succ 0}{\rm minimize} \:\dfrac{1}{n}\displaystyle\sum_{i=1}^{n}\by_{i}^{H}\bR^{-1}\by_{i} + \log|\bR|
\end{array},
\end{equation}
where $Band-Toep$ is used to denote the set of banded Toeplitz matrices. {Like in~\eqref{Opt:1}}, the above problem can be cast in the following equivalent form
\begin{equation}\label{bttemp}
\begin{array}{ll}
 \underset{{\bR\in Band-Toep, \bX}}{\rm minimize} \: \trace(\bX) + \log|\bR|\\
\hspace{6mm}{\rm subject\ to}\quad \: \left(\begin{array}{c c}
\bX&\bY^H\\
\bY&\bR
\end{array}\right) \succeq \mathbf{0}
\end{array}.
\end{equation}
Hence, (\ref{bttemp}) is handled via MM framework solving the following surrogate minimization problem
\vspace{1cm}
\begin{equation}\label{Bt}
\begin{array}{ll}
\underset{{\bE}}{\rm minimize} \: \|\bE - \bB\|_{F}^{2}\\
{\rm subject\ to}\:\: \bE + \bD \succeq \mathbf{0}\\
\hspace{17.5mm}\bE = \text{diag}(\bX,\bR)\: \textrm{with } \bR \textrm{ being a}\\
\hspace{17.5mm} \textrm{banded Toeplitz matrix}
\end{array}
\end{equation}
The above problem involves two convex sets: the set defined by the LMI $\bE +\bD \succeq \mathbf{0}$ and the set of block diagonal matrices where the second block has a banded Toeplitz structure with bandwidth $b$. Consequently, Dykstra's projection algorithm or POCS can be used to solve Problem (\ref{Bt}). The projection of a matrix onto the LMI set can be calculated as discussed earlier in Subsection \ref{sec:3c}. The projection of a matrix $\hat{\bPsi} {= \left(\begin{array}{c c}
	\hat{\bPsi}_{11}&\hat{\bPsi}_{12}\\
	\hat{\bPsi}^H_{12}&\hat{\bPsi}_{22}
\end{array}\right)}$ onto the set of block diagonal matrices with the second banded Toeplitz block can be obtained as follows. The first diagonal block is {the} same as $\hat{\bPsi}_{11}$ and the second diagonal block is constructed {by averaging the entries of the main and the first $b$ upper-}diagonals of the matrix $\hat{\bPsi}_{22}$ {and computing the corresponding Toeplitz matrix}~\cite{projbt}.
\subsubsection{MLE of {block-Toeplitz or} Toeplitz-block-Toeplitz covariance matrix}\label{subsection:ATOM2_TBT}
In space-time adaptive processing radar applications, the covariance matrix exhibits {a block-Toeplitz (BT) or} a Toeplitz-block-Toeplitz (TBT) structure. An example of a {BT}-structured covariance matrix with $p$ blocks is shown below
\begin{equation}\label{eq:BT_matrix}
	\bR=
	{\begin{bmatrix}
			\bR_{0} & \bR_{1} & \dots  & \bR_{p-1} \\
			\bR^{H}_{1} & \bR_{0}  & \dots  & \bR_{p-2} \\
			\vdots & \ddots & \ddots & \vdots \\
			\bR^{H}_{p-1} & \dots & \bR^{H}_{1} & \bR_{0}
	\end{bmatrix}}.
\end{equation}
{When each block exhibit a Toeplitz structure, then $\bR$ is TBT~\cite{tbt, tbt2}}.

The MLE problem of {a BT or a} TBT covariance matrix is formulated as
\begin{equation}\label{TBTtemp}
 \underset{\bR  \in {BT (TBT)}, \bR \succ 0}{\rm minimize} \:\dfrac{1}{n}\displaystyle\sum_{i=1}^{n}\by_{i}^{H}\bR^{-1}\by_{i} + \log|\bR|,
\end{equation}
where the notation {$BT (TBT)$} is used to indicate the set of {$BT (TBT)$} matrices. {A feasible solution to Problem~\eqref{TBTtemp} can be} obtained by solving at any given step the following surrogate optimization problem
\begin{equation}\label{TBT}
\begin{array}{ll}
\underset{{\bE}}{\rm minimize} \:~~ \|\bE - \bB\|_{F}^{2}\\
{\rm subject\ to}\:\: \bE +\bD \succeq \mathbf{0}\\
\hspace{17.5mm}{\bE\: \textrm{is a block diagonal matrix with}}\\
\hspace{17.5mm}{ \textrm{the second diagonal {BT (TBT)} block}}
\end{array}.
\end{equation}
Problem (\ref{TBT}) exhibits two constraints - $1$) a LMI constraint and $2$) a structural constraint - where the optimization variable $\bE$ is confined to be a block diagonal matrix with the second block having a {BT (TBT)} structure. Since both the constraints are convex, Dykstra's projection or POCS can be applied to solve Problem (\ref{TBT}). The projection of a matrix onto the LMI set can be calculated as discussed earlier in Section \ref{sec:3} B. The projection of a given matrix $\bar{\bPsi}$ onto the set of matrices whose second diagonal block has the {BT (TBT)} constraint can be obtained as follows. For the first diagonal block, the submatrix $\bar{\bPsi}_{11}$ is directly used. Then, the second diagonal block is obtained following {two (three)} steps. {First,} $p$ matrices are obtained by averaging the (upper-right) diagonal blocks of the matrix $\bar{\bPsi}_{22}$. Then, {only for TBT,} each of the $p$ matrices are projected onto the Toeplitz set as described in subsection~\ref{sec:3c}. {Finally, the resulting matrix is constructed according to~\eqref{eq:BT_matrix}.}
\section{CRB calculation}\label{sec:crlb}
In this section, the {CRB} is derived for {the estimation of} {Toeplitz structured covariance matrix (the interesting reader may refer to Appendix C of the supplementary material with reference to the CRBs of Banded Toeplitz, BT, and TBT covariance model}). The CRB provides a lower bound on the variance of any unbiased estimator \cite{kay}. To proceed further, let {$\bth$} represent the real value vector parametrizing a given covariance matrix structure of interest. 
Then, the CRB is the inverse of the Fisher Information matrix (FIM) whose $(i,k)^{th}$ element is
\begin{equation}
\begin{array}{ll}
[\bF]_{i,k} = \textbf{E}\left[\frac{\partial^{2} \log {\bar{f}(\bR)}}{\partial \theta_{i}\partial\theta_{k}}\right]
\end{array},
\end{equation}
where {$\frac{\partial \log {\bar{f}(\bR)}}{\partial \theta_{i}}$ denotes the partial derivative of $\log {\bar{f}(\bR)}$ w.r.t. $\theta_{i}$, with $\theta_{i}$ the $i$-th element of $\bth$}.
 {Due to the Gaussian assumption, the} $(i,k)^{th}$ element of the FIM can be computed using the \emph{Slepian–Bangs
formula} \cite{stoica}
\begin{equation}\label{FIM}
[\bF]_{i,k} = n\textrm{Tr}\left(\bR^{-1}\frac{\partial\bR}{\partial \theta_{i}}\bR^{-1}\frac{\partial\bR}{\partial \theta_{k}}\right).
\end{equation}
{In the following subsection, {the FIM is derived} for the Toeplitz covariance structure}.

\subsection{Toeplitz matrix}\label{sec:4a}
As the entries of the TSC matrix are completely characterized by its first row, i.e., $[r_{1}, r_{2},\cdots r_{m}]^{T}$, the covariance matrix $\bR \in \mathbb{H}^{m \times m}$ can be parameterized by $\bth = [r_{1}, \Re(r_{2}),\cdots\Re(r_{m}),\Im(r_{2}),...,\Im(r_{m}) ]^{T} \in \mathbb{R}^ {2m-1}$ where $\Re(r_{i})$ and $\Im(r_{i})$ denotes the real and imaginary parts of $r_{i}$, respectively. Then, the covariance matrix $\bR$ can be expressed in terms of $\bth$ and basis matrices $\bB^{\rm{Toep}}_{g}$ (defined as in (\ref{basis})), $g=1,2,\cdots,m$ \cite{em2}
\begin{equation}\label{toep}
    \begin{array}{ll}
    \bR = \displaystyle\sum_{g=1}^{m}\theta_{g}{\Re}(\bB^{\rm{Toep}}_{g}) + j { \displaystyle\sum_{g=m+1}^{2m-1}\theta_{g}{\Im}(\bB^{\rm{Toep}}_{g-m+1})}
    \end{array}.
\end{equation}
The $(i,k)^{th}$ element of the matrix $\bB^{\rm{Toep}}_{g}$ is given as
\begin{equation}\label{basis}
\begin{array}{ll}
    [\bB^{\rm{Toep}}_{g}]_{i,k}=
\begin{cases}
    {1+j}& i-k=g-1=0\\
    1+j & k-i=g-1\neq0\\
    1-j&i-k=g-1\neq 0\\
    0 &\rm{otherwise}
\end{cases}
\end{array}.
\end{equation}
Using (\ref{toep}), $\frac{\partial\bR}{\partial \theta_{i}}$ can be  obtained as
\[
    \frac{\partial\bR}{\partial \theta_{i}}=
\begin{cases}
    \Re(\bB^{\rm{Toep}}_{i}) & 1\leq i \leq m\\
    j\Im(\bB^{\rm{Toep}}_{i-m+1})& m+1 \leq i \leq 2m-1
\end{cases}
\]
Substituting $\frac{\partial\bR}{\partial \theta_{i}}$ in (\ref{FIM}), yields the FIM for Toeplitz covariance matrix.

\section{Numerical Simulations}\label{sec:4}
In this section, the performance of the proposed covariance matrix estimators ATOM1 and ATOM2 is {numerically} analyzed in comparison with {the following} state-of-the-art algorithms: {EM-based \cite{em1, fuhrmann1990estimation}}, MELT \cite{melt}, the SCM, {and the FB estimators~\cite{vantrees4}}. {First, a convergence analysis of the derived methods is {provided}, also in comparison with the aforementioned counterparts. Then, the estimation capabilities are analyzed in three different scenarios, using the MSE as performance metric, {defined} as\footnote{{In the following,~\eqref{mse} is computed via Monte Carlo techniques}.}
\begin{equation}\label{mse}
	\textrm{MSE} =      \textbf{E}\left[\left\|\bth - \hat{\bth}\right\|^2\right] ,
\end{equation}
where $\hat{\bth}$ indicates the estimate of the unknown $\bth$}{, obtained according to one of the aforementioned strategies.}
{First of all, the covariance matrix is assumed to share the Toeplitz structure. Then, the banded Toeplitz, the BT, and the TBT constraints are considered. The CRB{-based benchmark}, computed as $\text{CRB} = \trace(\bF^{-1})$, {is reported too}, whereby, for each {case study}, the FIM is appropriately derived, {see} Section~\ref{sec:crlb}.}

Furthermore, {assuming a typical radar signal processing scenario,} the performance is {also} evaluated in terms of average achievable SINR {by an adaptive spatial filter}.

\begin{figure}[t] \centering
	\includegraphics[width=0.80\linewidth]{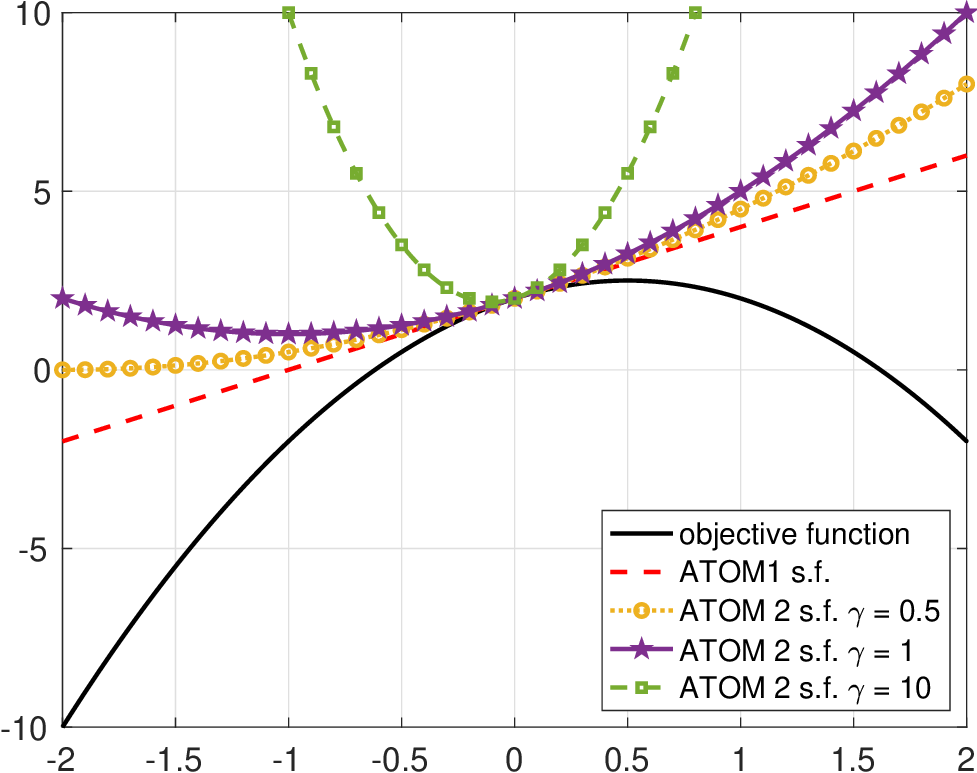}
	\caption{{A notional representation of the objective function of Problem~\eqref{Opt:1} and the corresponding s.f. of ATOM1 and ATOM2, with the latter employing $\gamma \in \{0.5, 1, 10\}$, for a one-dimensional optimization problem.}}
	\label{fig:atom_opt}
\end{figure}

{It is also worth reporting that, in the aforementioned scenarios, ATOM1 and ATOM2 procedures are initialized using the FB estimate $\bR_{FB}$, {projected onto the set of Toeplitz matrices}. Moreover, for the execution of ATOM2, the parameter $\gamma$ is updated adaptively in each outer-loop iteration according to the following law\footnote{{As to the adaptive ATOM2 surrogate construction stage, it has been empirically shown that the updating rule~\eqref{eq:gamma}, with $\gamma_0= 10^{-4}$ and $k_1 = 5$, provides satisfactory performance in all the scenarios; therefore, unless otherwise stated, ATOM2 s.f. (and the subsequent processing) is constructed using~\eqref{eq:gamma} with the aforementioned values.}}
	\begin{equation}\label{eq:gamma}
		\gamma = \gamma_0 (t \log{t+{k_1}})^2.
	\end{equation}
To illustrate the role of $\gamma$ in the optimization process performed by ATOM2, a notional representation of the objective function (conceptually depicted as a one-dimensional curve {and corresponding to a specific portion of a restriction of the multivariate objective}) and the s.f. of ATOM1 and ATOM2, is reported in Fig.~\ref{fig:atom_opt}. 
Remarkably, the value of $\gamma$ affects the trade-off between performance and convergence speed of ATOM2. Indeed, while a smaller $\gamma$ leads to {a} better performance (ATOM2 s.f. approaches the ATOM1 one as $\gamma \rightarrow 0$), it demands more inner-loop iterations to achieve convergence, {due to the almost singular resulting metric}. On the other hand, a larger $\gamma$ reduces the overall computational cost, but introduces a growth in the approximation error. However, as the outer-loop iterations increase, the approximation error of the ATOM2 s.f. w.r.t. the objective function decreases as the {updated point} becomes closer and closer to {a local} minimum {at which the sequence is ``converging''}. That said, slowly increasing $\gamma$ with the number of iterations allows to speed-up its computational burden without decreasing its performance.}

\subsection{Assessment of iterative algorithms convergence for on-grid and off-grid frequencies}\label{sec:5a}
In this simulation, the convergence of ATOM1 and ATOM2 (whose inner-loop was implemented via Dykstra's algorithm) is assessed in comparison with MELT and EM algorithms. To this end, {each} data snapshot $\by_{k}\in \mathbb{C}^m$ is modeled as
\begin{equation}\label{data}
    \by_{k}= \bR^{\frac{1}{2}}\bn_{k},\; k=1,2, \cdots, n
\end{equation}
{where $\bn_{k} \in \mathbb{C}^m, \;k=1,\dots, n$ are independent and identically distributed zero-mean circularly symmetric Gaussian random vectors with unit mean square value.}

\begin{figure}[t] \centering
	\subfloat[]{\label{fig:negLL_obj_ON_GRID_a}		\includegraphics[width=0.48\linewidth]{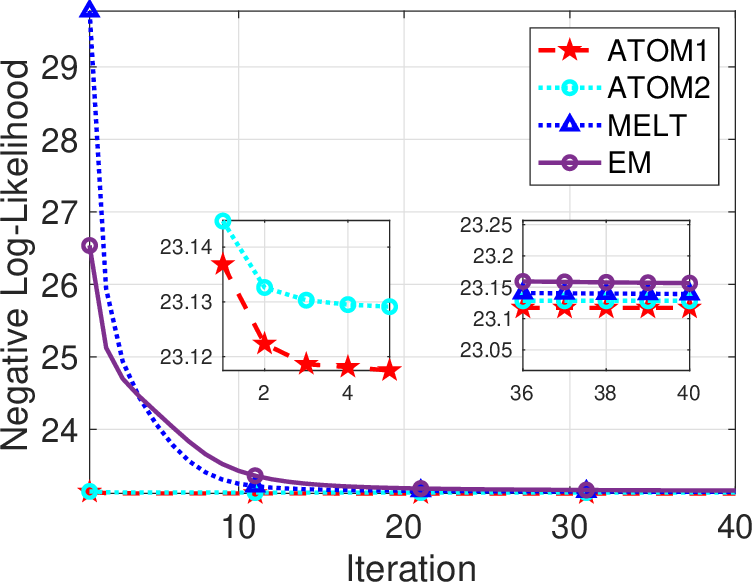}} \hfill
	\subfloat[]{\label{fig:negLL_obj_ON_GRID_b}			\includegraphics[width=0.49\linewidth]{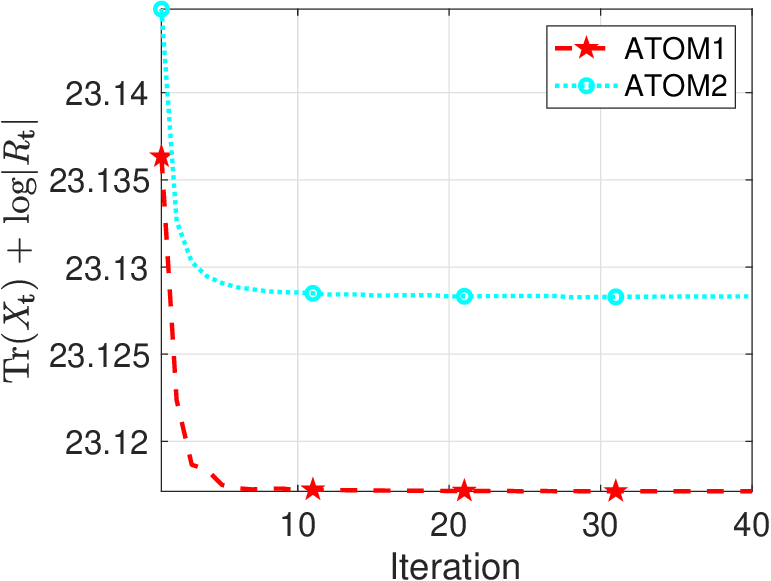}}
	\caption{{Negative log-likelihood~\eqref{eq:obj_function} and the objective function of~\eqref{Opt:1} vs. outer-iterations for $m=6$, $n=20$, and on-grid frequencies scenario}.}
	\label{fig:negLL_obj_ON_GRID}
\end{figure}

{Two different experimental setups are considered, assuming $m=6$ and $n=20$. In the former,} the true underlying Toeplitz covariance matrix $\bR$ is constructed by choosing the $2$-{nd}, $3$-rd, $5$-th, $7$-th, $8$-th and the $11$-th {column} of the DFT matrix with $L=2m-1$ in (\ref{ce}), corresponding {to the} frequencies $[0.5712, 1.1424, 2.2848, 3.4272, 3.9984, 5.7120]$ rad, {and as powers $[p_1, \dots, p_6]^\mathrm{T} = [3, 6, 4, 1,  7, 5]^\mathrm{T}$}, respectively. {Figs. \subref*{fig:negLL_obj_ON_GRID_a} and \subref*{fig:negLL_obj_ON_GRID_b} show the negative log likelihood~\eqref{eq:obj_function} and the objective function of problem~\eqref{Opt:1}} versus the number of iterations, respectively. It can be seen that all the algorithms numerically {improve the negative log-likelihood as the number of iterations increases} and {almost} converge to the same {value}, {with negligible differences}. {Moreover, Fig.~\subref*{fig:negLL_obj_ON_GRID_b} indicates that the proposed algorithms monotonically decrease the problem objective function, which is expected since they optimize~\eqref{Opt:1} using the MM framework.}

\begin{figure}[t] \centering
	\subfloat[]{\label{fig:negLL_obj_OFF_GRID_a}
		\includegraphics[width=0.48\linewidth]{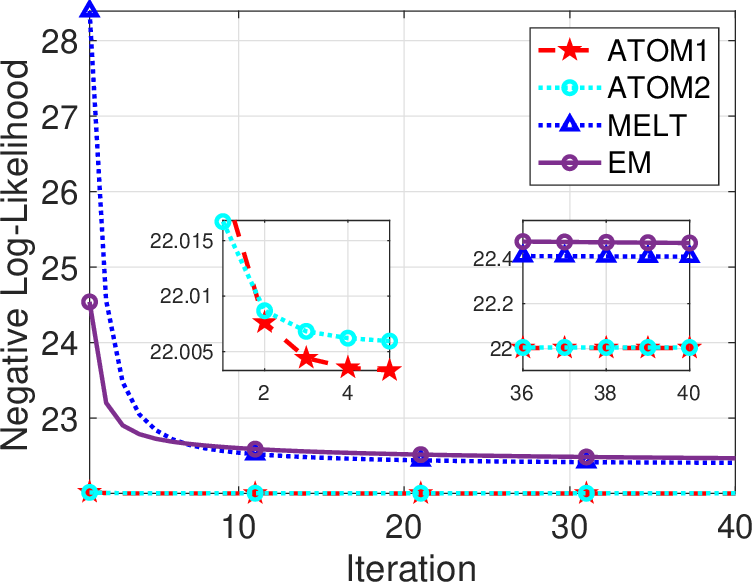}}\hfill
	\subfloat[]{\label{fig:negLL_obj_OFF_GRID_b}
		\includegraphics[width=0.49\linewidth]{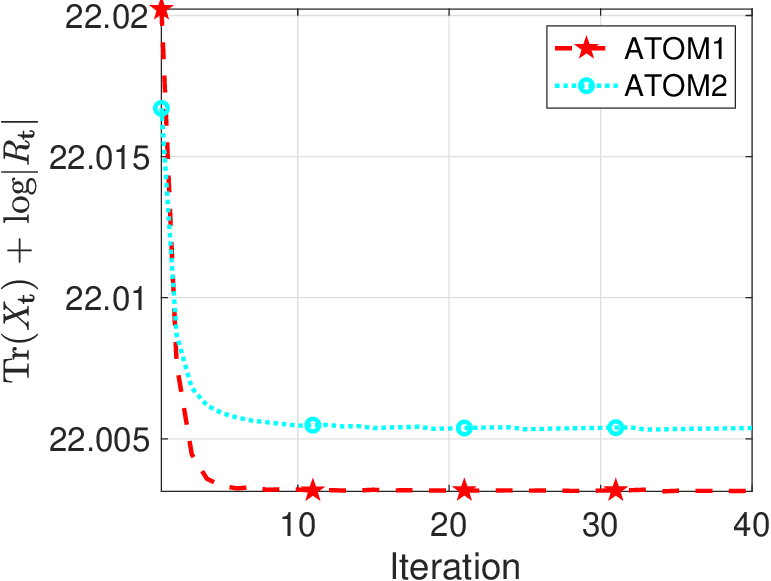}}
	\caption{{Negative log-likelihood~\eqref{eq:obj_function} and the objective function of~\eqref{Opt:1} vs. outer-iterations for $m=6$, $n=20$, and off-grid frequencies scenario}.}
	\label{fig:negLL_obj_OFF_GRID}
\end{figure}

In the {other} experimental setup, the true underlying Toeplitz covariance matrix is constructed such that {two} of the frequencies are not on the Fourier grid. {Therefore, the same parameters used} in case study 1 {are considered}, with the exception that the Fourier frequencies {$0.5712$ rad and $3.9984$ rad are replaced with $0.5$ rad and $5.3$ rad, respectively}. {For the case study at hand, the negative log-likelihood~\eqref{eq:obj_function} and the objective function of~\eqref{Opt:1} are reported in Figs.~\subref*{fig:negLL_obj_OFF_GRID_a} and~\subref*{fig:negLL_obj_OFF_GRID_b}  versus the number of iterations, respectively.} {Inspection of Fig.~\subref*{fig:negLL_obj_OFF_GRID_a} reveals} that while MELT and EM converge to a value of {$\approx 22.4$}, ATOM1 and ATOM2 converge to $22$. Therefore, when {two} of the frequencies do not lie on the Fourier grid, the state-of-the-art iterative algorithms converge to a larger value of the negative log-likelihood {than} the proposed {methods}. This is {due to the fact that} unlike the counterparts, the proposed algorithms estimate the Toeplitz covariance matrix without reparametrizing it via the CE technique and thus {they are able to} cover the whole set of Toeplitz covariance matrices. {Furthermore, remarks similar to those made for the on-grid case hold true with reference to the results depicted in Fig.~\subref*{fig:negLL_obj_OFF_GRID_b}.}

{In the following, the mean computational time\footnote{{The simulation has been executed using MATLAB R2020b on a desktop computer equipped with an Intel i5 processor and 16 GB of RAM.}} (averaged over 1000 Monte Carlo trials) of the proposed techniques and the counterparts is examined. As case studies, four different values of $m$ are considered, i.e., $m \in \{4, 8, 16, 32\}$. Moreover, the} data samples $\by_{k}$ are generated as (\ref{data}) {using} $n=4m$ samples, with $\bm{R} = \bm{T} + \bm{I}$. The Toeplitz covariance matrix $\bT$ is generated {assuming 3 equal power sources, i.e., with $p = [5, 5, 5]$, whose frequencies are randomly selected (at each trial) such that two of them lie on the Fourier grid of the DFT matrix, with $L=2m-1$, whereas the third one is drawn from a uniform distribution over $[0, 2\pi]$}. The iterative algorithms have been run until the following condition is met\footnote{{For the execution of EM and MELT procedures, the exit condition is set as $f(\bR_{t-1})-f(\bR_{t}) \leq 10^{-4}$.}}
{\begin{equation}
	p(\bR_{t-1}, \bX_{t-1})-p(\bR_{t}, \bX_{t}) \leq 10^{-4}
\end{equation}
with $p(\bR, \bX) = \trace(\bX) + \log|\bR|$ the objective function of problem~\eqref{Opt:1},}
or until the maximum number of iterations (set equal to $1000$) is reached. 
{The average computational time of the different algorithms {(possibly with different values of the hyperparameters)} are reported in} Table~\ref{table:run_time}.
 The results show that {ATOM2 has, in general, a longer execution time than ATOM1}. This is because the inner-loop of {ATOM2} {(based on} Dykstra's algorithm) requires an {higher} number of iterations and hence a {longer} run time to converge than ATOM1 inner-loop (implemented via ADMM), {and similar to those of EM/MELT} {when $\gamma_0$ is small, {where the distance is minimized in a metric space is ill defined more and more.} However, when $\gamma_0 = 10^{-1}$, the run times of ATOM1 and ATOM2 are comparable {and similar to those of MELT and EM}.} {Interestingly, Table~\ref{table:MSE} pinpoints that, for  $\gamma_0$ sufficiently small, i.e., $10^{-4}$, ATOM2 is generally able to reach MSE values smaller than ATOM1, reasonably to its adaptive \textit{step-size} strategy~\eqref{eq:gamma}, which allows it to provide better quality estimates than ATOM1 as the outer-loop iteration increases.} It can also be seen that {EM} has the least computational time {(at large values of $m$)}. Nevertheless, {as shown in Table~\ref{table:MSE},} although the proposed algorithms have a {slight} longer computational time, {the obtained estimates are superior, in terms of MSE, to those provided by MELT and EM}. 
 
 {Interestingly, as the data dimension increases, the resulting average MSE values reached by the ATOM2 using different $\gamma_0$ parameters becomes closer and closer. Therefore, for a sufficient larger data size, i.e., $m\ge32$, $\gamma_0 = 10^{-1}$ represents an appropriate choice for ATOM2 implementation, as it offers a good performance with a reduced computational burden.}

\subsection{MSE vs $n$ for Toeplitz covariance matrix}\label{sec:5b}
\begin{figure}[t]\centering
		\subfloat[]{\label{fig:MSE_a}	\includegraphics[width=0.80\linewidth]{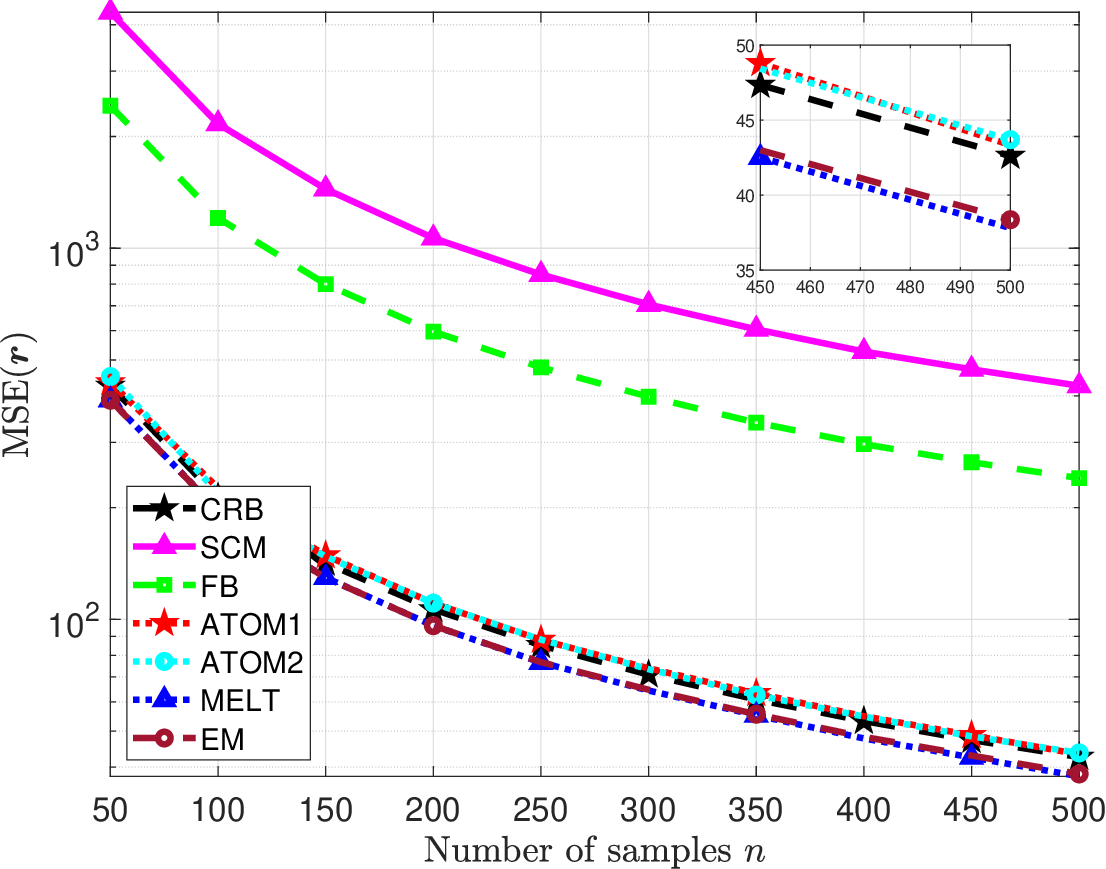}}\\
	\subfloat[]{\label{fig:MSE_b}		\includegraphics[width=0.80\linewidth]{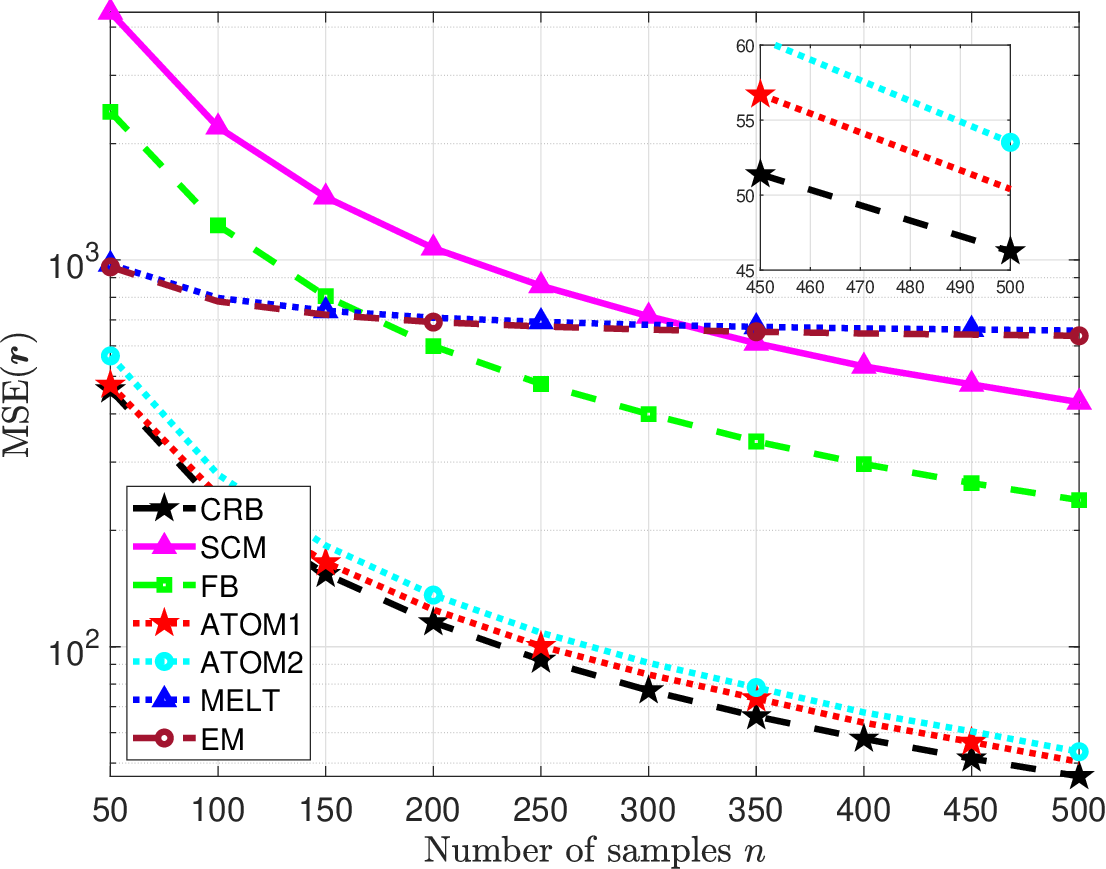} }
	\caption{MSE vs. number of samples $n$ for Toeplitz covariance matrix. a) on-grid frequencies; b) off-grid frequencies.}
	\label{fig:MSE}
\end{figure}

{For this {case studies}, it is assumed $m= 15$ and the number of samples $n$ ranging between $50$ and $500$ in steps of $50$. The data} $\by_{k} \in \mathbb{C}^{15}$ are again simulated {according to~\eqref{data}}. 
{Precisely, two different experiments are considered whereby the} true Toeplitz covariance matrix is generated using on-grid\footnote{The frequencies used in the first experiment are: $[0.2167, 0.6500, 1.0833, 1.3, 1.5166, 1.9500, 2.3833, 2.8166, 3.2499,$ $3.6832 4.1166, 4.5499, 4.9832, 5.4165, 5.8499]$ rad. Their corresponding {powers} increase linearly from $1$ to $15$ with a unit step.} and off-grid frequencies\footnote{{For the off-grid simulation, the frequencies $[1.3, 2.8166, 4.9832,5.8499]$ rad are replaced with $[1.25, 3.01, 5.20, 5.8]$ rad, respectively.}}, respectively. 
{The resulting MSE, computed over 1000 Monte Carlo trials, are illustrated in Fig.~\ref{fig:MSE}.}
{Inspection of the curves depicted in Fig.~\subref*{fig:MSE_a}} shows that, {regardless of the number of samples $n$,} in the first experiment ATOM1 and ATOM2 {almost reach the CRB}, whereas EM and MELT {yield} {a slightly better performance, resulting in a deviation from the CRB. This can be explained observing that the derived CRB does not exploit the information that the frequencies lie on-grid.} Fig.~\subref*{fig:MSE_b} highlight that in the second experiment, ATOM1 {attain the best performance, with results quite close to the CRB and slightly better than ATOM2, with a limited gap between the corresponding curves}. Furthermore, MELT and EM exhibit similar MSE {values which seem to saturate as $n$ increases. The performance behavior of Fig.~\subref*{fig:MSE_b} stems from the observation that}, unlike MELT and EM, ATOM1 and ATOM2 {are gridless methods, delivering the same performance regardless of the sources frequencies}.

\begin{table*}[t]
	\centering
	\caption {Comparison of the average run time (in seconds) of the iterative algorithms.}
	\label{table:run_time}
	\resizebox{1\linewidth}{!}{
		\begin{tabular}{c a c c c c c}
			\hline\hline
			\rowcolor{white}
			\textbf{Dimension $m$}& \textbf{ATOM1}& \textbf{ATOM2 ($\gamma_0 = 10^{-4}$)}  & \textbf{ATOM2 ($\gamma_0 = 10^{-2}$)}  & \textbf{ATOM2 ($\gamma_0 = 10^{-1}$)}  &\textbf{MELT}\cite{melt}&\textbf{EM}\cite{em1} \\
			\hline
			4&\cellcolor{White}0.028&1.309&0.047&\cellcolor{Gray}0.014&0.051&0.026\\
			8&0.032&1.503&0.164&0.055&0.071&0.035\\
			16&0.163&6.912&0.522&0.166&0.162&0.081\\
			32&0.473&9.569&2.484&0.825&0.663&0.348\\
			\hline
			\hline
	\end{tabular}}
\end{table*}

\begin{table*}[t]
	\centering
	\caption {Comparison of the average MSE of the iterative algorithms.}
	\label{table:MSE}
	\resizebox{1\linewidth}{!}{
		\begin{tabular}{c c a c c c c}
			\hline\hline
			\rowcolor{white}
			\textbf{Dimension $m$}& \textbf{ATOM1}& \textbf{ATOM2 ($\gamma_0 = 10^{-4}$)}  & \textbf{ATOM2 ($\gamma_0 = 10^{-2}$)}  & \textbf{ATOM2 ($\gamma_0 = 10^{-1}$)}  &\textbf{MELT}\cite{melt}&\textbf{EM}\cite{em1} \\
			\hline
			4&42.48&38.12&45.08&47.88&45.04&44.64\\
			8&22.80&19.92&23.04&23.92&82.48&82.32\\
			16&30.88&26.40&32.80&35.36&93.60&91.36\\
			32&20.16&20.16&20.16&20.48&112.96&107.21\\
			\hline
			\hline
	\end{tabular}}
\end{table*}

\subsection{MSE vs $n$ for banded Toeplitz covariance matrix}\label{sec:5c}
{This subsection analyzes the performance in the case of covariance matrix belonging to the set of banded Toeplitz matrices. In particular, the same simulation setup {as in} Section~\ref{sec:5b} is considered, but enforcing the underlying covariance matrix to have a bandwidth $b=6$. To this end, $\bR$ is} constructed by alternately projecting a  random Hermitian matrix onto the set of banded Toeplitz matrices and the set of PSD matrices. 
{Moreover, for this study case, ATOM2 is implemented according to the procedure described in Section~\ref{subsection:ATOM2_bandedT}, namely explicitly including the banded Toeplitz structure in the constraint set.}

\begin{figure}[t]\centering
	\includegraphics[width=0.80\linewidth]{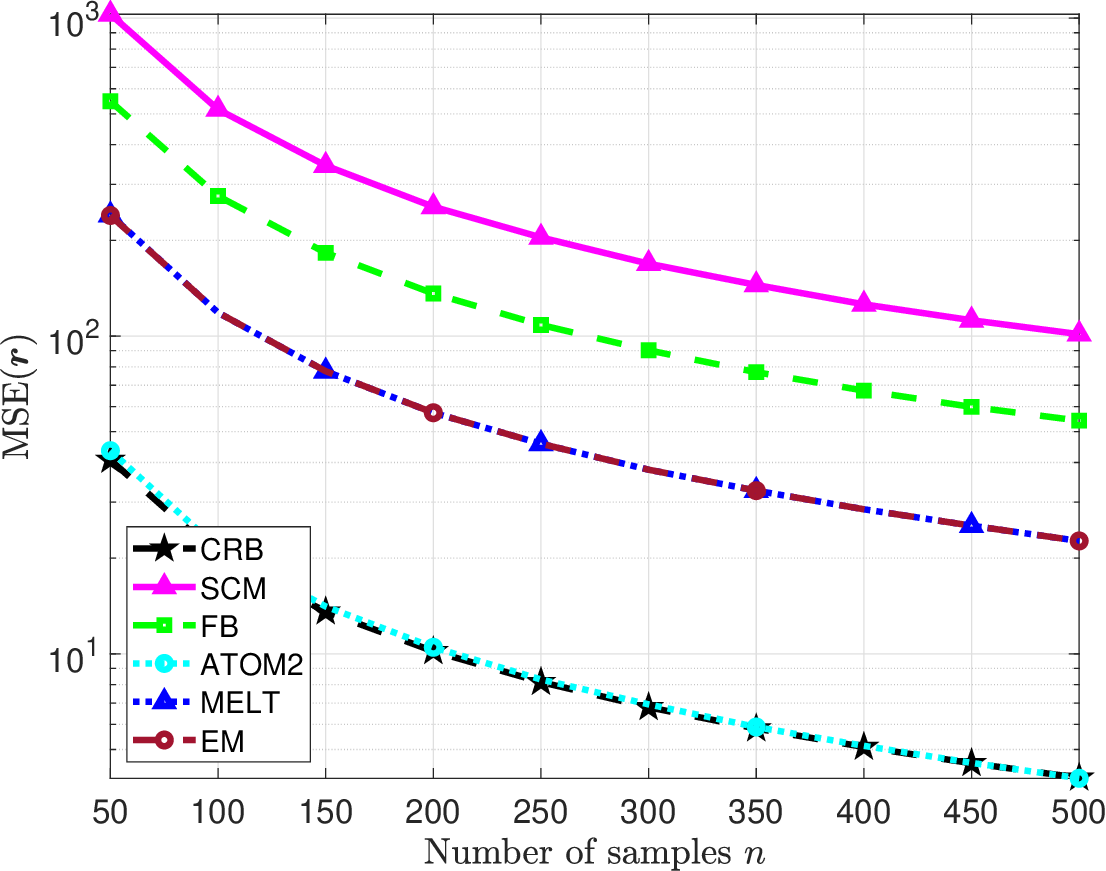}
	\caption{MSE vs. number of samples $n$ for banded Toeplitz covariance matrix.}
	\label{fig:MSE_bandedT}
\end{figure}

Fig.~\ref{fig:MSE_bandedT} highlights that the {bespoke implementation of ATOM2 delivers the best performance, with MSE values really close to the CRB. Furthermore, MELT  and EM share the same performance with a noticeable gap w.r.t. ATOM2, which is expected since the aforementioned algorithms do not leverage the banded structure of the covariance matrix.}

\begin{figure}[t] \centering
	\includegraphics[width=0.80\linewidth]{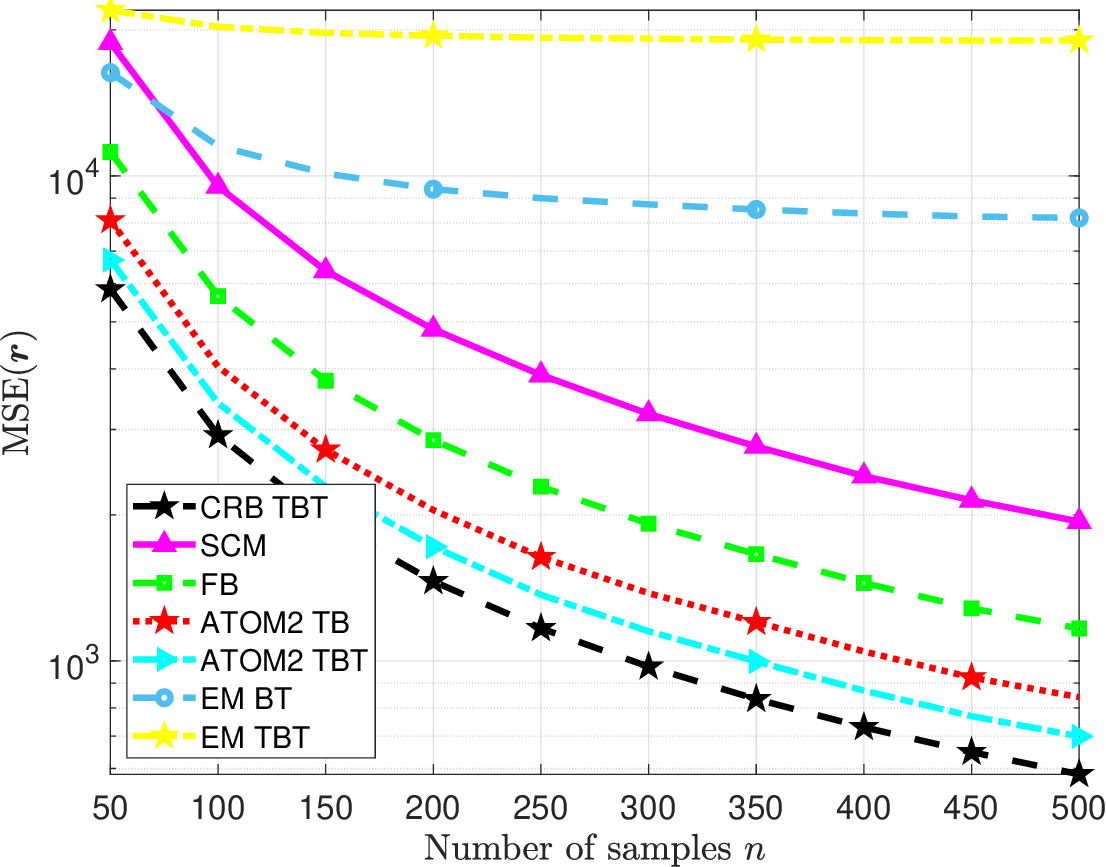}
	\caption{MSE vs. number of samples $n$ for TBT covariance matrix.}
	\label{fig:MSE_TBT}
\end{figure}

\subsection{MSE vs $n$ for {BT (TBT)} covariance matrix}\label{sec:5d}
{Here, the capabilities of ATOM2 are analyzed in the context of covariance matrix with TBT structure. To this end, assuming $m=16$ and $p=4$ blocks (each having block-size $l=4$), the covariance matrix is {modeled} as $\bR = \bT_1 \otimes \bT_1$, where $\bT_1 \in \mathbb{C}^{l \times l}$ is a Toeplitz matrix constructed as in subsection~\ref{sec:5a}, with frequencies $[0.6, 1.4, 3.2, 5.1]$ rad and powers $[3,6,4,1]$. Thus, each data snapshot $\by_{k}$ is {drawn} according to~\eqref{data}.
The resulting MSE values (averaged over 1000 Monte Carlo trials) are displayed in Figure~\ref{fig:MSE_TBT} versus the number of {snapshots}. Specifically, {the performance of} both the BT and the TBT extension of ATOM2 (described in Section~\ref{subsection:ATOM2_TBT}) are reported and compared with the CRB {(see Appendix C reported in the supplementary material to this paper)} {as well as with two EM-based estimators, tailored respectively for BT/TBT} covariance matrix~\cite{fuhrmann1990estimation}.
Inspection of the results reveals that ATOM2 TBT uniformly achieves the least MSE, with ATOM2 BT ranking second}. As previously highlighted, the superior performance of the proposed method stems from the design criterion which does not require reparametrizing the covariance matrix using the CE.

\subsection{Radar Application}\label{sec:5f}
In this subsection, the performance of the covariance estimation algorithms is evaluated with reference to the {average} achievable SINR in {adaptive} radar spatial processing context. To this end, let us consider a radar system {equipped with a} uniform linear array with $m=6$ sensors, pointing toward the boresight direction. The {inter-element distance} between each sensor is set equal to $d=\lambda/2$, where $\lambda$ is the radar operating wavelength. 

{For this simulation scenario,} the interference covariance matrix is {modeled} as $\bR = \bR_{s} + \sigma_{a}^{2}\bI$ where $\sigma_{a}^{2}$ is the power level of the white disturbance noise {(assumed without loss of generality equal to {0 dB})} and $\bR_{s}$ is {given by $\bR_{s} =  {\sum\limits_{l=1}^{J}} \sigma_l^2\: \bs(\phi_l) \bs(\phi_l)^H$, where $J$ is the number of uncorrelated narrow-band jammers and, for the $l$-th jammer, 
\begin{eqnarray}\label{eq:steering_vector}
	\bs(\phi_l) = {\frac{1}{\sqrt{m}}}[1, e^{j \frac{2\pi}{\lambda} d \sin(\phi_l)}, \dots, e^{j (m-1)  \frac{2\pi}{\lambda} d \sin(\phi_l)}]^{\mathrm{T}}
\end{eqnarray}
is the steering vector in its direction-of-arrival $\phi_l$, and $\sigma^2_l$ the corresponding interferer power}.

{The capabilities of the estimation methods are analyzed by means of the average SINR, computed as}
\begin{equation}\label{eq:SINR_avg}
    {\rm{SINR}}_{\rm{avg}}= \dfrac{1}{K}\displaystyle\sum_{i=1}^{K}\dfrac{|\hat{\bw_{i}}^{H}\bs(\theta)|^{2}}{\hat{\bw}_{i}^{H}\bR\hat{\bw_{i}}},
\end{equation}
where {$K=500$} is the number of Monte-Carlo trials and ${\hat{\bw}_{i}} = {\hat{\bR}_{i}}^{-1}\bs(\theta)$ is the estimate of the optimal weight vector for {adaptive} spatial processing with ${\hat{\bR}_{i}}$ the estimate of the interference{-plus-noise} covariance matrix for the $i$-th trial, {computed either via the sample covariance matrix or enforcing the Toeplitz structure in the covariance matrix and employing the estimators ATOM1, ATOM2, EM, and MELT.}

{More precisely, $J=2$ jammers, with powers $\sigma_{1}^{2}= 30$ dB and $\sigma_{2}^{2}= 20$ dB, respectively, impinging on the array from $\theta_{1}=9.8^{\circ}$ and $\theta_{2}=-8.8^{\circ}$, {is} considered. As comparison terms, the optimum SINR, i.e., $ {\rm{SINR}}_{\rm{OPT}} = \bs(\theta)^H{{\bR}}^{-1}\bs(\theta)$ and {the performance of} the Sample Matrix Inversion (SMI) beamformer, are included {too}.}

\begin{figure*}[t] \centering
	\subfloat[]{		\includegraphics[width=0.32\linewidth]{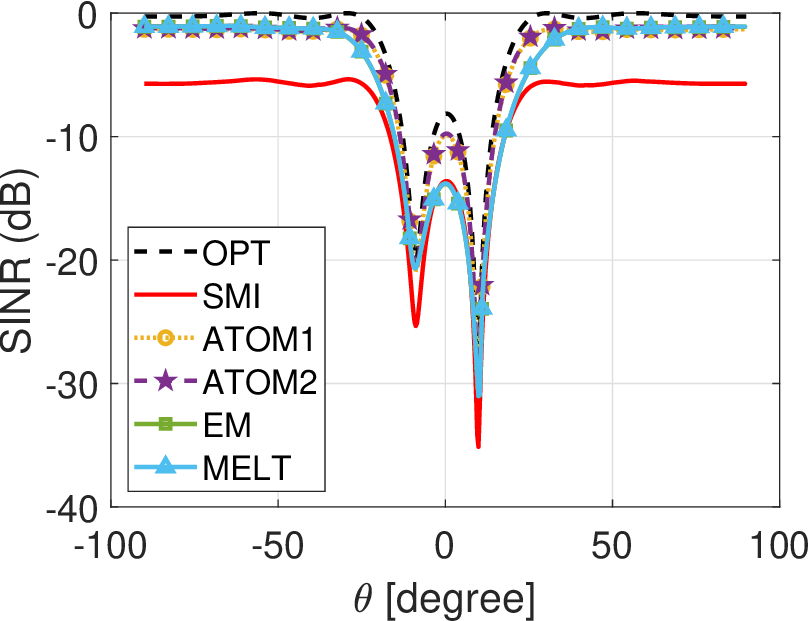}}\hfil
	\subfloat[]{		\includegraphics[width=0.32\linewidth]{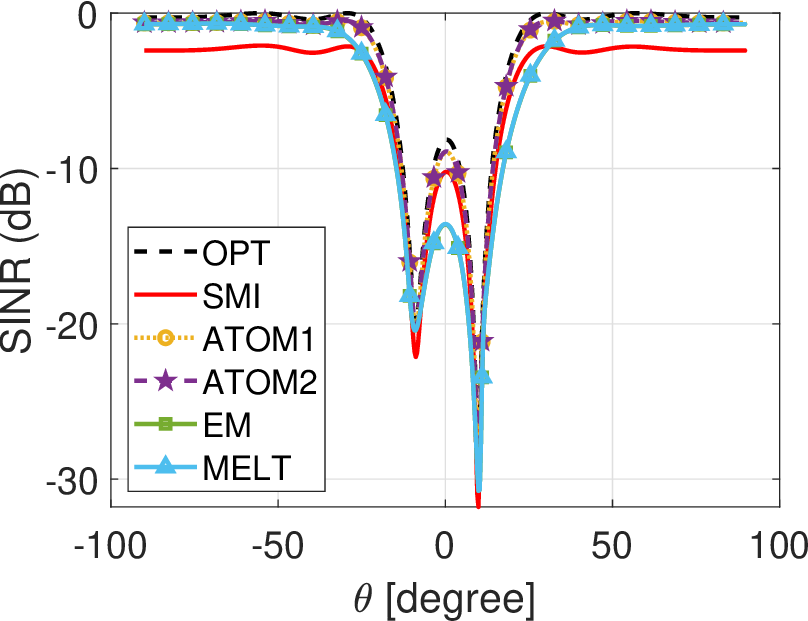} }\hfil
	\subfloat[]{		\includegraphics[width=0.32\linewidth]{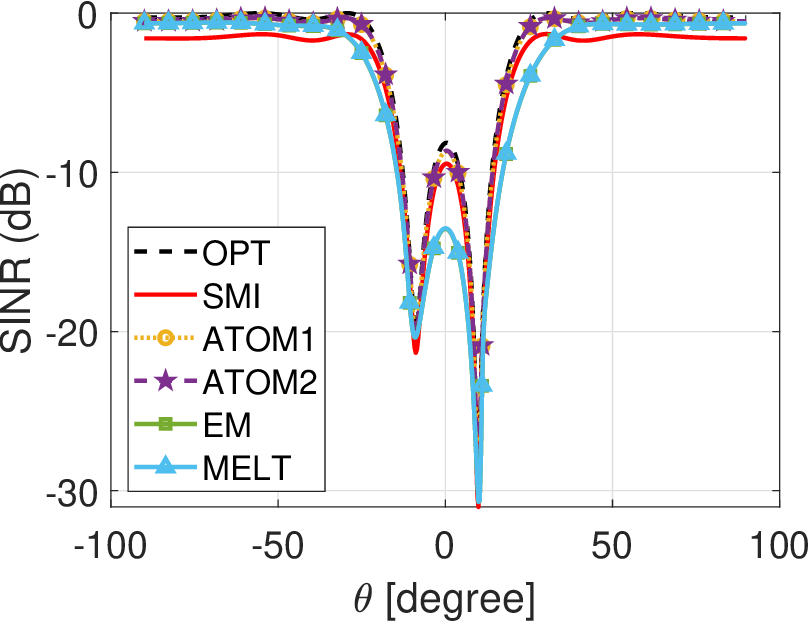} }
\caption{Average SINR vs $\theta$ {in the presence of two jammers, assuming $m=6$ and a) $n=m $ b) $n=2m$, and c) $n=3m$}.}
\label{fig:SINR_radar}
\end{figure*}

{The} average SINR {versus} {$\theta \in \mathcal{T}$, with $\mathcal{T = }[-\pi/2, \pi/2]$ discretized with 500 equally-spaced points,} is shown in {Fig.~\ref{fig:SINR_radar}, for $n\in\{m, 2m, 3m\}$}. Inspection of the plots highlights that as the number of samples $n$ increases, {the results achieved by} ATOM1 and ATOM2 gets {closer and closer} to the {optimum, yielding superior performance w.r.t. the counterparts}.

\section{Conclusion}\label{sec:5}
In this paper, {the} MLE {problem for} TSC matrices {has been addressed}. Precisely, {by reformulating appropriately the MLE optimization problem and} leveraging the MM framework, two iterative algorithms ATOM1 and ATOM2 {have been} developed. Both inherit the key properties of MM i.e., they monotonically decrease the underlying cost function with guaranteed convergence to a stationary point of the equivalent MLE problem. Subsequently, ATOM2 {has been} extended to handle covariance matrix MLE forcing other Toeplitz-related structures, such as banded Toeplitz, {BT,} and TBT. Simulation results {have indicated} that the proposed algorithms can perform better than some state-of-the-art techniques in terms of MSE and the SINR metrics. \\
Some of the possible future research directions are now outlined. {In particular, ATOM2 could be further extended to include the cases of low rank TSC, with the rank assumed either known or unknown at the design stage, as well as covariance matrix with an upper bound to the condition number.}
Another possible extension of the proposed technique could be MLE of a Toeplitz covariance matrix assuming a compound Gaussian distribution for the underlining data which has a significant application in low-grazing angle target detection \cite{appp,gini}. {Moreover, acceleration methods inspired for instance by the SQUAREd iterative Methods (SQUAREM)~\cite{JSSv092i07} could be investigated. Finally, the design of sub-optimal optimization strategies (e.g., based on the gradient projection method) with an improved computational burden (a valuable feature for real-time applications) is definitely worth to be pursued}.

\section*{Appendix A}\label{appendixA}
\section*{Proof of equivalence between (8) and (10)}
\red{Let $\bR^\star$ be an optimal solution to (8), then $(\bX^\star, \bR^\star)$, with $\bX^\star= \bY^H \bR^{\star-1} \bY$, is feasible for (10) and the two problems have the same objective values. This means that
	\begin{equation}\label{eq:A}
		v(8) \ge v(10),
	\end{equation}
	where $v(\cdot)$ indicates the optimal value of the corresponding optimization problem.}

\red{
	Moreover, for any fixed $\bR_1 \succ 0$, concentrating the objective function of (10) with respect to $\bX$ (which is tantamount to placing $\bX= \bY^H \bR_1^{-1} \bY$),  it follows that the concentrated optimization problem is
	\begin{equation}\label{eq:app_2}
		\underset{\bR_1 \succeq 0}{\rm minimize} \: \trace(\bR_{FB} \bR_1^{-1}) + \log|\bR_1|,
	\end{equation}
	due to Schur complement Theorem and the monotonicity of the trace operator with respect to generalized matrix inequality ``$\succeq$''.
	Finally, being by assumption (8) solvable, any minimizer of~\eqref{eq:app_2} satisfies $\bR_1^\star \succ 0$ with a corresponding optimal solution to (10) given by $(\bR_1^\star, \bY^H \bR_1^{\star-1} \bY)$. This implies that 
	\begin{equation}\label{eq:B}
		v(8)\le v(10).
	\end{equation}
	Capitalizing on~\eqref{eq:A} and ~\eqref{eq:B} as well as the above considerations, it follows that $v(8)=v(10)$ and given an optimal solution $(\bR_1^\star,\bX_1^\star)$ to (10), $\bR_1^\star$ is also optimal to (8) and viceversa, given an optimal solution $\bR^\star$ to (8) $(\bX^\star, \bR^\star)$ is an optimal point to (10).}

\section*{Appendix B}\label{appendixC}
\section*{Proof of Theorem 3.2}
To begin with, let us denote by \red{$h(\bE|\bE_{t})$ either} the objective function involved in the surrogate optimization problem \red{of ATOM1 (12) or ATOM2 (15)}, where $\bE = \text{diag}(\bX, \bR)$. This function, regardless of the method, satisfies the following two inequalities
\begin{equation}\label{d1}
	\red{h}(\bE_t|\bE_{t}) = \red{l}(\bE_{t})
\end{equation}
\begin{equation}\label{d2}
	\red{h}(\bE_{t+1}|\bE_{t}) \geq \red{l}(\bE_{t+1})
\end{equation}
where $\red{l}(\bE)= \textrm{Tr}(\bX) + \log|\bR|$. Leveraging the above inequalities, it follows that
\begin{equation}\label{d3}
	\red{l}(\bE_{t+1})  \overset{(a)} \leq \red{h}(\bE_{t+1}|\bE_{t})  \overset{(b)} \leq \red{h}(\bE_{t}|\bE_{t})  \overset{(c)}= \red{l}(\bE_{t})
\end{equation}
In (\ref{d3}), the inequality (a) and equality (c) stem from (\ref{d2}) and (\ref{d1}), respectively; besides, the inequality (b) is obtained by exploiting the fact that ATOM1 and ATOM2 globally solve the corresponding convex surrogate optimization problem. Therefore, (\ref{d3}) implies that the sequence of objective value of Problem (16) generated by the proposed algorithms is monotonically decreasing , i.e.,
\begin{equation}\label{d4}
	\red{l}(\bE_{0}) \geq \red{l}(\bE_{1}) \geq \red{l}(\bE_{2}) \geq \cdots
\end{equation}
Next, let us denote by $\bZ$ a cluster point to $\{\bE_{t}\}$ and let $\{\bE_{r_{t}}\}$ be a subsequence  of $\{\bE_{t}\}$ converging to $\bZ$. Then, from (\ref{d1}), (\ref{d2}), and (\ref{d4})
\begin{equation}
	\begin{array}{ll}
		\red{h}\left(\bE_{r_{t+1}}|\bE_{r_{t+1}}\right)= \red{l}\left(\bE_{t_{j+1}}\right) \leq \red{l}\left(\bE_{r_{t}+1}\right)\\\hspace{5mm}\leq
		\red{h}\left(\bE_{r_{t}+1}|\bE_{r_{t}}\right)\leq \red{h}\left(\bE|\bE_{r_{t}}\right), \forall\,\, {\mbox{feasible}}\,\,\bE.
	\end{array}
\end{equation}
Thus, letting $t \rightarrow \infty$
\begin{equation}
	\red{h}(\bZ|\bZ) \leq \red{h}(\bE|\bZ),
\end{equation}
which implies that  $\red{h}'(\bZ|\bZ;\bD) \geq 0$ where $\red{h}'(\cdot|\bZ;\bD)$ is the directional derivative of the surrogate function at point $\bZ$ in a feasible direction $\bD$. Finally, by Proposition 1 in \cite{conv}, the surrogate function $\red{h}(\bE|\bZ)$ and  the objective function $\red{l}(\cdot)$ have the same first order behavior at $\bZ$. Therefore, $\red{h}'(\bZ|\bZ;\bD) \geq 0$ implies that  $\red{l}'(\bZ; \bD) \geq 0$. Hence, $\bZ$ is a stationary
point of the objective function $\red{l}(\bE)$.

\section*{Appendix C}
\section*{CRB of Banded Toeplitz, BT, and TBT covariance model}
Herein, the CRB of Banded Toeplitz, BT, and TBT covariance model are provided.

\subsection{Banded Toeplitz matrix}\label{sec:4b}
In the case of banded Toeplitz matrix with bandwidth $b$, the first row of the covariance matrix $\bR  \in \mathbb{H}^{m \times m}$ has only $b+1$ non-zero terms. Therefore, $\bR$ can be parameterized via $\bth = [r_{1}, \Re(r_{2}),\cdots\Re(r_{b+1}),\Im(r_{2}),...,\Im(r_{b+1}) ]^{T} \in \mathbb{R}^ {2b+1}$. Besides $\bR$ can be expressed in terms of basis matrices $\bB^{\rm{Toep}}_{g}$ and real coefficients $\bth$
\begin{equation}
	\begin{array}{ll}
		\bR = \displaystyle\sum_{g=1}^{b+1}\theta_{g}{\Re}(\bB^{\rm{Toep}}_{g}) + j \displaystyle\sum_{g=b+2}^{2b+1}\theta_{g}{\Im}(\bB^{Toep}_{g-b})
	\end{array}
\end{equation}
and consequently
\[
\frac{\partial\bR}{\partial \theta_{i}}=
\begin{cases}
	{\Re}(\bB^{\rm{Toep}}_{i}) & 1\leq i \leq b+1\\
	j{\Im}(\bB^{\rm{Toep}}_{i-b})& b+2\leq i \leq 2b+1
\end{cases}.
\]
Substituting $\frac{\partial\bR}{\partial \theta_{i}}$ in (34), yields the FIM for banded Toeplitz covariance matrix.

\subsection{Toeplitz-block-Toeplitz matrix}\label{sec:4c}
{Before proceeding further, it is worth noting that a TBT matrix composed of $p$ blocks of size $l$ can be parameterized by the vector $\bth = [\bth_0^T, \bth_1^T, \dots, \bth_{P-1}^T]^{{T}} \in \mathbb{R}^{2 l -1 + (p-1)(4l-2)}$ whereby $\bth_0 = [r_{0,1}, \Re({r_{0,2}}), \dots, \Re({r_{0,l}}), \Im({r_{0,2}}), \dots, \Im({r_{0,l}})]^{{T}} \in \mathbb{R}^{2l-1}$ and $\bth_p = [\Re({r_{p,1}}), \dots, \Re({r_{p,l}}), \Im({r_{p,1}}), \dots, \Im({r_{p,l}}),\\ \qquad \Re({c_{p,2}}), \dots, \Re({r_{p,l}}), \Im({r_{p,2}}), \dots, \Im({r_{p,l}})]^{{T}}\in \mathbb{R}^{4l-2}, \;p=1,\dots, P-1$, with $r_{p,n}$ and $c_{p,n}$ the $n$-th row and $n$-th column of $\bR_{p}$, respectively.}
{Indeed}, the TBT covariance matrix can be expressed as
\begin{eqnarray}
	\bR^{\rm{TBT}} = \bC_{0}\otimes\bR_{0} +\displaystyle\sum_{w=1}^{{p-1}} \left(\left(\bC_{w}\otimes{\bR^{H}_{w}}\right) +\left(\bC_{w}^{T}\otimes{\bR_{w}}\right)\right),
\end{eqnarray}
where
{\begin{equation}
		\bR_0 = 
		\sum_{g=1}^{l}\theta_{0,g}{\Re}(\bB^{\rm{Toep}}_{g}) + j \sum_{g={l+1}}^{2l-1} \theta_{0,g}{\Im}(\bB^{Toep}_{g-l+1})
\end{equation}}
{and, for $w=1,\dots, p-1$,
	\begin{equation}
		\begin{aligned}
			\bR_{w} = & \sum_{g=1}^{l}[\theta_{w,g} + j\theta_{w,g+l} ]{\Re}(\bD_g) \\
			& + \sum_{g=2l+1}^{3l-1}[\theta_{w,g} + j\theta_{w,g+l-1} ]{\Im}(\bD_{g-2l+1})
		\end{aligned}
	\end{equation}
	with $\theta_{w,g}$ the $g$-th element of $\bth_w$, $\bD_g = \bB^{\rm{Toep}}_{g}$ as long as $g = 1$ and $1/2 ((\bB^{\rm{Toep}}_{g})^T + j (\bB^{\rm{Toep}}_{g})^T)$ elsewhere}, whereas the $(i,k)^{th}$ element of the matrix $\bC_{w} \in \mathbb{R}^{l\times l}$ is given by
\[
[\bC_{w}]_{i,k}=
\begin{cases}
	1& i-k=w\\
	0 & \rm{otherwise}
\end{cases}.
\]

{That said,
	$\frac{\partial\bR^{\rm{TBT}}}{\partial \theta_{w,g}}$ is given by
	\begin{equation*}
		\resizebox{0.19\hsize}{!}{$\frac{\partial\bR^{\rm{TBT}}}{\partial \theta_{w,g}}=$ }
		\resizebox{0.81\hsize}{!}{$
			\begin{cases}
				\bC_{0} \otimes \Re(\bB^{\rm{Toep}}_{g}) & 1\leq g\leq{l}, w=0\\
				\bC_{0} \otimes j\Im(\bB^{\rm{Toep}}_{g-l+1})& {l}+1\leq g \leq 2l-1, w=0\\
				\bC_{w} \otimes\Re(\bD_{g})^T \\ \;\;+ \bC_{w}^{T} \otimes{\Re}(\bD_{g})  & 1\leq g \leq l, w > 0 \\
				\bC_{w} \otimes(-j)\Re(\bD_{g-l})^T \\ \;\;+ \bC_{w}^{T} \otimes{j\Re}(\bD_{g-l})  & l+1\leq g \leq 2l, w > 0 \\
				\bC_{w} \otimes\Im(\bD_{g-2l+1})^T \\ \;\;+ \bC_{w}^{T} \otimes{\Im}(\bD_{g-2l+1})  & 2l+1\leq g \leq 3l-1, w > 0 \\
				\bC_{w} \otimes(-j)\Im(\bD_{g-3l+2})^T \\ \;\;+ \bC_{w}^{T} \otimes{j\Im}(\bD_{g-3l+2})  & 3l\leq g \leq 4l-2, w > 0\\
			\end{cases}$ }
	\end{equation*}
	which, employed in (34), yields} the FIM for TBT covariance matrix.

\bibliographystyle{IEEEtran}
\bibliography{refs}
\end{document}